\documentclass[12pt, a4paper]{article} 

\usepackage[left=3cm,right=3cm,top=3cm,bottom=3cm]{geometry}

\usepackage[utf8]{inputenc}
\usepackage[english]{babel}

\usepackage{bbm} 
\usepackage{amsmath}
\usepackage{amsfonts}
\usepackage{amssymb}
\usepackage{amsthm}
\usepackage{dsfont}

\usepackage{multirow}
\usepackage{adjustbox}
\usepackage{booktabs}

\usepackage{caption}
\usepackage{subcaption}

\usepackage{rotating}

\usepackage{natbib}
\usepackage{appendix}

\newcommand{\I}{\mathcal{I}}
\newcommand{\E}{\mathrm{E}}
\newcommand{\Var}{\mathrm{Var}}

\newcommand{\Cov}{\mathrm{Cov}} 
 
\newcommand{\argmin}{\operatornamewithlimits{argmin}}

\newtheorem{proposition}{Proposition}

\newtheorem{corollary}{Corollary}
\theoremstyle{definition}
\newtheorem{definition}{Definition}
\newtheorem*{example}{Example}

\begin{document}
	
\title{The Murphy Decomposition and the Calibration-Resolution Principle: A New Perspective on Forecast Evaluation
	 \thanks{
		\textbf{Acknowledgements:} I thank Uwe Hassler and Patrick Schmidt for helpful comments. Earlier versions of this paper were presented at DAGStat 2019 in Munich and the International Symposium on Forecasting 2019 in Thessaloniki.}}

\author{Marc-Oliver Pohle\thanks{pohle@econ.uni-frankfurt.de, Address: Campus Westend, RuW
		Building, Theodor-W.-Adorno-Platz 4, 60323 Frankfurt, Germany} \\
	Goethe University Frankfurt}	 

\maketitle

\begin{abstract}

I provide a unifying perspective on forecast evaluation, characterizing accurate forecasts of all types, from simple point to complete probabilistic forecasts, in terms of two fundamental underlying properties, autocalibration and resolution, which can be interpreted as describing a lack of systematic mistakes and a high information content. This ``calibration-resolution principle'' gives a new insight into the nature of forecasting and generalizes the famous sharpness principle by \cite{gneiting2007sharpness} from probabilistic to all types of forecasts. It amongst others exposes the shortcomings of several widely used forecast evaluation methods. The principle is based on a fully general version of the Murphy decomposition of loss functions, which I provide. Special cases of this decomposition are well-known and widely used in meteorology.

Besides using the decomposition in this new theoretical way, after having introduced it and the underlying properties in a proper theoretical framework, accompanied by an illustrative example, I also employ it in its classical sense as a forecast evaluation method as the meteorologists do: As such, it unveils the driving forces behind forecast errors and complements classical forecast evaluation methods. I discuss estimation of the decomposition via kernel regression and then apply it to popular economic forecasts. Analysis of mean forecasts from the US Survey of Professional Forecasters and quantile forecasts derived from Bank of England fan charts indeed yield interesting new insights and highlight the potential of the method.

\end{abstract}

\newline


\section{Introduction}

High-quality forecasts are the basis for sound decision-making. Probabilistic forecasts, i.e.\ forecasts of the full distribution of the variable of interest, give the complete picture and have become increasingly popular in recent years (see e.g.\ \cite{gneiting2014}), but are often difficult to obtain. Point forecasts transport a limited amount of information, but are easier to construct and are still the most widely issued type of forecast in many disciplines. Interval forecasts provide a good compromise between the two. No matter of what type the forecasts are, forecast evaluation plays a crucial role in the provision of good forecasts, detecting the strengths and weaknesses of forecasters or forecasting methods and guiding them towards possible improvements.

The standard method for evaluating the most common types of forecasts, namely mean and median forecasts, is the measurement of their accuracy via a suitable loss function. In contrast to that, traditional methods for the evaluation of quantile or interval forecasts are tests of unconditional or conditional exceedance or coverage (see e.g.\ \cite{christoffersen1998}), which are tests of optimality relative to some information set.\footnote{Optimality tests are also common for other types of forecasts, (see e.g.\ \citet[chapter 15]{elliott2016}), for example mean forecasts, but certainly not the main method of evaluation.}
For probabilistic forecasts, a widely used traditional method of evaluation is the assessment of uniformity of the probability integral transform [PIT] (see e.g.\ \cite{dawid1984} or \cite{diebold1998}). 

Taking a general perspective on forecast evaluation, in seminal work, \cite{gneiting2013} and \cite{gneiting2007proper} advocate the evaluation of all types of forecasts via suitable loss functions, namely consistent scoring functions for point forecasts and proper scoring rules for probabilistic forecasts. In other words, they put forward the assessment of forecast accuracy as the base of forecast evaluation for all types of forecasts and stress the importance of the use of loss functions appropriate for the respective type.

Focusing on the evaluation of probabilistic forecasts, in a related paper \cite{gneiting2007sharpness} criticize the sole use of the PIT for evaluating such forecasts and conjecture the so-called sharpness principle, stating that good probabilistic forecasts should maximize sharpness, i.e.\ concentration of the predictive distribution, subject to calibration, and thereby characterizing accurate forecasts in terms of fundamental underlying properties. \cite{tsyplakov2011} later proves the principle after specifying that the required notion of calibration is autocalibration. However, while considering the principle as very interesting from a theoretical point of view, he puts into question its practical usefulness as its applicability relies on the forecasts being perfectly autocalibrated. Nevertheless, the sharpness principle has been very influential for the theory and practice of the evaluation of probabilistic forecasts over the last years.

Accompanying this principle by diagnostic tools for the evaluation of calibration and sharpness and again advocating the use of proper scoring rules, \cite{gneiting2007sharpness} strive to set up a general framework for probabilistic forecasts in the tradition of \cite{murphy1987}, who proposed a general framework for the evaluation of point forecasts. \cite{murphy1987} emphasized the analysis of features of the joint distribution of forecasts and observations to complement predictive accuracy assessment to gain deeper insights into the strengths and weaknesses of forecasting methods and to finally improve them.\footnote{Note that they had mainly probability forecasts of binary events and categorical variables and mean forecasts in mind and not the more complicated quantile, interval or distributional forecasts, for which \cite{gneiting2013} and \cite{gneiting2007proper} had to stress years later that the base of forecast evaluation should be the measurement of predictive accuracy.} The article shaped the meteorological forecast evaluation (or verification as it is called there) literature, which still uses partly quite different methods than statisticians or econometricians, putting a larger emphasis on properties of the joint distribution of forecasts and observations or on conditional distributions of one given the other (see e.g.\ \cite{wilks2011} or \cite{jolliffe2012} for overviews).   

In this work I on the one hand provide a generalization of the sharpness principle by characterizing the accuracy of all types of forecasts, including point forecasts, by underlying theoretical properties and discuss implications of this. On the other hand, I show that the use and the advancement of forecast evaluation methods from the meteorological literature, which emphasize the use of such underlying properties in the spirit of \cite{murphy1987}, are beneficial also for other disciplines by applications to popular economic forecasts.

Both of this is based on a fully general version of the Murphy decomposition from meteorology, which decomposes the forecasting loss into three components: Uncertainty, representing the variation in the outcome variable, resolution, which describes the ability of the forecasts to distinguish between different outcomes, and mis(auto)calibration. (Auto)calibration refers to the statistical consistency between forecasts and observations. Resolution can also be interpreted as measuring the informational content of the forecasts and autocalibration as freedom from systematic forecasting mistakes. The decomposition was originally proposed by \cite{murphy1973} for the case of a probability forecast for a binary event using the Brier score as a loss function. 
This original version of the decomposition has been vividly used in meteorology as a forecast evaluation method (see again e.g.\ \cite{wilks2011} or \cite{jolliffe2012}) and has also spread to other disciplines like psychology (see e.g. \cite{slovic1977} or \cite{koriat1980}). In the statistical and econometric forecast evaluation literature, the decomposition is not very well known, see \cite{diebold1996} and \cite{elliott2016} for two exceptions in the form of a handbook article and a book respectively. To the best of my knowledge the only two papers from statistics or econometrics that apply the decomposition (to economic data) are \cite{galbraith2012} and \cite{lahiri2013}. Despite presenting very interesting results, these papers are limited to probability forecasts for binary events and it is certainly of interest to analyze decompositions for other types of forecasts like mean forecasts of continuous variables, quantile or interval forecasts or even full distributional forecasts.  The meteorological literature has progressed in this direction: Focussing on a specific loss function, \cite{murphy1996} introduces the decomposition for the case of mean forecasts and quadratic loss, \cite{hersbach2000} introduces it for distributional forecasts and the continuous ranked probability score, \cite{todter2012} introduce it for variants of the logarithmic score and \cite{bentzien2014} introduce it for quantile forecasts and the quantile score. Aiming at more general classes of loss functions, \cite{broecker2009} treats proper scoring rules for discrete distributional forecasts and \cite{ehm2017} treat consistent scoring functions. I provide the fully general version of the decomposition, which captures all types of forecasts that can be evaluated by a consistent scoring function or a proper scoring rule.

While the Murphy decomposition has only been used empirically so far, i.e.\ as a forecast evaluation method, in the first part of the paper I use it in a theoretical way. I point out that it yields a generalization of the influential sharpness principle by \cite{gneiting2007sharpness} to all types of forecasts, which I call the ``calibration-resolution principle'' and show that the sharpness principle indeed arises as a special case for autocalibrated probabilistic forecasts. The new principle thus on the one hand solves the above-mentioned problem that the sharpness principle is only applicable for autocalibrated forecasts, which could be even more severe as \cite{tsyplakov2011} thought due to a possible trade-off between maximizing resolution and minimizing miscalibration that a forecaster could often face in practice as I argue. On the other hand point forecasts are captured by the new principle, too. The calibration-resolution principle characterizes accurate forecasts in terms of two fundamental underlying properties, gives us a better understanding of the nature of forecasting and a new way to think about what it means to construct accurate forecasts, namely jointly maximizing information content and minimizing systematic mistakes. 

Demonstrating that the principle can foster our understanding of forecast evaluation in general, I consider some widely-used traditional forecast evaluation methods, e.g.\ the PIT or the assessment of conditional and unconditional exceedance or coverage, some of which have already been criticized in the literature (see e.g. \cite{hamill2001}, \cite{gneiting2007sharpness}, \cite{holzmann2014} or \cite{nolde2017}). From the angle of the calibration-resolution principle I point out that all of these methods assess some form of calibration, but do not capture the information content of the forecasts, resulting in an incomplete evaluation if used as a sole evaluation method. Furthermore, I discuss the relationship between optimality testing and calibration.

Before introducing the calibration-resolution principle and drawing conclusions from it in the third section of this paper, I set up the basic framework and carefully introduce autocalibration and resolution in the second section: I pay special attention to explicitly modelling the time series and information structure and to introducing the concept of resolution, which is hardly known outside of the meteorological literature. Furthermore, I provide two justifications for interpreting resolution as a measure of information content: I firstly show that if the information contained in a forecast is larger than the one contained in another forecast, resolution is higher for that forecast. Secondly, I derive a generalization of the total variance formula, where resolution equals the term which generalizes the explained part of the variance. To illustrate the ideas behind autocalibration and especially resolution and to make clear which dimensions of forecast quality they capture and how they complement each other, I use a theoretical example comprising several stylized versions of forecasters, which builds on an example from \cite{gneiting2007sharpness}. This example also illustrates the Murphy decomposition and enriches the discussion of the sharpness and the calibration-resolution principle.

In the fourth section, I use the Murphy decomposition in the (at least for meteorologists) classical way, as a forecast evaluation method,  and highlight its potential as a complement to standard forecast evaluation techniques usually applied in statistics and econometrics due to the new insights it yields by applying it to popular examples of economic forecasts. The decomposition unveils the driving forces behind forecast errors, showing whether they are mainly caused by systematic mistakes or a lack of information content and where improvements are needed. It also monitors the fall in information content with rising forecast horizon and uncovers the forecast horizon up to which the forecasts contain usual information at all, giving hints on the limits of forecastability. 

At the beginning of the fourth section I discuss estimation of the decomposition terms, for which conditional functionals of the predictive distribution given the forecasts, e.g.\ conditional quantiles or the conditional expectation have to be estimated. I concentrate on point forecasts here as for probabilistic forecasts conditioning variables are high-dimensional and the estimation is in general problematic due to the curse of dimensionality. While the meteorologists use naive methods based on binning, I employ local linear kernel regressions, where the suitable consistent scoring functions should be used as the respective regression loss functions, e.g.\ for estimating conditional means least-squares kernel regression and for estimating conditional quantiles quantile regression are the right choices. Note that \cite{galbraith2011} use local constant kernel regressions for estimating conditional probabilities of a binary event in a similar context.

Subsequently I apply the general form of the decomposition to economic forecasts for the first time. The applications are to mean forecasts of inflation and GDP growth from the US Survey of Professional Forecasters [SPF] and to quantile forecasts obtained from the probabilistic inflation and GDP growth forecasts issued by the Bank of England [BoE]. The former show no signs of miscalibration and I can trace back their huge drop in performance from nowcasts to one-quarter-ahead forecasts to a lack of resolution. Despite the lack of resolution, they clearly outperform the classical benchmark in the form of autoregressive [AR] models, hinting to the inherent difficulty of this forecasting problem. 

The BoE's probabilistic forecasts, which I evaluate over a range of quantiles, show a diverse pattern in terms of the decomposition with multiple interesting insights: For both variables at most quantiles there is considerable resolution for the shorter forecast horizons, which gradually declines towards zero and often a rise in miscalibration from then on. Furthermore, substantial differences between the quantiles can be observed, where the forecasts are better at the lower parts of the distribution and deteriorate towards the upper parts. However, the inflation forecasts contain much more information content and (as is analyzed in \cite{pohle2020benchmark}) clearly outperform relevant benchmarks, while the GDP growth forecasts are sometimes outperformed by a benchmark, namely the quantile autoregressive [QAR] model, indicating potential for improvement. 

The fifth section sketches ideas for future research and forecast evaluation practice and concludes. The appendix contains in four parts proofs, calculations for the illustrative example used throughout the paper, details on robustness checks and benchmark forecasts and details on the probabilistic forecasts issued by the BoE and how to calculate quantiles from them.

\section{Fundamentals, Autocalibration and Resolution}

Even though forecast accuracy, characterized by closeness between forecasts and observations in terms of a suitable loss function, is the primary criterion for evaluating forecasts, many other properties that characterize good forecasts exist. Examining those properties may yield valuable additional insights and lead to improved and in turn more accurate forecasts as \cite{murphy1987} stress in their influential paper. Due to the influence of this paper on the meteorological forecast evaluation literature, such fundamental properties as (auto)calibration and resolution are widely used there, but hardly known in other disciplines. In this section I thus introduce these two properties, justify their use and illustrate them by a theoretical example. While they are very interesting for forecast evaluation on their own right, the next section demonstrates via the so-called calibration-resolution principle that they are intimately related to forecast accuracy, which makes them even more relevant. To set the stage for the rest of the paper I first introduce the basic setup, review relevant classes of loss functions and the concepts of entropy and divergence.

\subsection{Basic Setup}

In a time series or sequential prediction problem a forecaster tries to predict a variable of interest $Y$ over one or several forecasting horizons given certain information in every period. All other variables which may contain relevant information concerning $Y$ are summarized in the possibly very large vector $Z$. The forecaster's target may be to forecast the whole distribution of $Y$, then he is called a probabilistic forecaster, or some functional of it, e.g.\ the mean or a certain quantile. 

To make the sequential nature of the problem and the information structure explicit, consider the real-valued time series $\{ Y_t \}_{t \in \mathbb Z}$ and $\{ Z_t \}_{t \in \mathbb Z}$. Denote the $\sigma$-algebra containing all the information at time $t$ $\mathcal{F}_t$, i.e.\ $\mathcal{F}_t=\sigma(Y_t, Z_t, Y_{t-1}, Z_{t-1}, ...)$, the $\sigma$-algebra generated by the history of $\{ Y_t \}$ and $\{ Z_t \}$. A forecaster who tries to predict $h$ periods into the future at time $t-h$ has access to a certain information set $\mathcal{I}_{t-h} \subset \mathcal{F}_{t-h}$. $Y_t$ may be continuous or discrete (with the important binary special case) random variables.

A probabilistic forecaster trying to predict $Y_{t}$ with a forecast horizon $h$ issues a forecast distribution or predictive distribution $F^{pr}_{Y_{t}}$, which is his or her best guess of the true distribution of $Y_{t}$ based on the given information contained in $\mathcal{I}_{t-h}$, $F_{Y_{t}|{\mathcal{I}_{t-h}}}$. The goal of a forecaster in general is to predict a functional $T$ of $F_{Y_{t}|{\mathcal{I}_{t-h}}}$, which he or she does by reporting the respective functional of his or her forecast distribution $T(F^{pr}_{Y_t})$. Most often a point forecast is required, e.g.\ a mean forecast with target functional $\E[Y_{t}|\mathcal{I}_{t-h}]$ or a quantile forecast, where the functional is a certain quantile $F^{-1}_{Y_{t}|\mathcal{I}_{t-h}}(\tau)$, $\tau \in (0,1)$. The functionals could be set-valued as e.g.\ in the case of quantiles of discrete distributions, but to streamline the discussion we assume unique functionals, noting that non-unique ones can be easily accounted for. The forecast targets may also be prediction intervals or histogram-type forecasts, which lead to vector-valued functionals, or of course the whole distribution.\footnote{As mentioned in the introduction, probability forecasts of a binary event do also frequently appear. Here, $Y_t$ is a binary random variable and the goal is thus to forecast the parameter of a Bernoulli distribution, which is at the same time the expected value and characterizes the whole distribution, i.e.\ in this important special case the forecast is a point forecast and a distributional forecast at the same time.} In all these cases, I will use the notation $T(F^{pr}_{Y_{t}})$, no matter if $T$ maps to the real numbers, a vector of real numbers or (for a probabilistic forecaster) to a density function or back to the distribution function itself. Furthermore, to simplify notation, I denote the forecast, no matter of what type it is, as $X_{t}=T(F^{pr}_{Y_{t}})$. This leads to a sequence of forecasts $\{ X_t \}_{t \in \mathbb Z}$. The information contained in the forecast $X_{t}$ is represented by $\sigma(X_{t}) \subset \mathcal{I}_{t-h}$, the $\sigma$-algebra generated by it. 

Note that even though I talk about a time series prediction problem here, cross-sectional forecasting problems are of course nested as a special case. The theory and methods laid out in this paper can be fruitfully applied to such problems, even a very large vector of explanatory variables $Z_t$ poses no problems, but on the contrary could make the results obtained even more interesting.

\subsection{Loss Functions}

The goal of forecast evaluation is to assess the quality of forecasts, which may have been obtained by any forecasting method, be it model-based or not. An indispensable step in the forecast evaluation process is measuring forecast accuracy via a suitable loss function. Loss functions also play a fundamental role in this paper. Thus, I review the relevant classes now.

In the statistical forecast evaluation literature, a loss function $s(x_t,y_t)$ mapping a realized forecast-observation pair $(x_t,y_t)$ onto the real numbers is called a scoring function if $x_t$ is a point forecast, i.e.\ if the underlying functional $T$ is one-dimensional, and a scoring rule if $x_t$ is a distributional or an interval forecast (see e.g.\ \cite{gneiting2007proper}, \cite{gneiting2011} and \cite{gneiting2014}). I define the loss functions as nonnegative and negatively oriented as is common in the literature. The forecaster's aim thus is to minimize $\E[s(X_{t},Y_{t})|\mathcal{I}_{t-h}]$, the expected score conditional on the given information set $\mathcal{I}_{t-h}$ by choosing the forecast $X_{t}$. A scoring function is called consistent or a scoring rule proper for the functional $T$ if this functional of the true conditional distribution of the outcome variable minimizes the expected score given the available information, formally if
$$\E[s(T(F_{Y_{t}|{\mathcal{I}_{t-h}}}),Y_{t})|\mathcal{I}_{t-h}] \leq \E[s(X_{t},Y_{t})|\mathcal{I}_{t-h}]$$
for all forecasts $X_t$, which are constructed on the basis of this information, i.e.\ which are $\mathcal{I}_{t-h}$-measurable, and strictly consistent or proper if equality holds only for $X_t = T(F_{Y_{t}|{\mathcal{I}_{t-h}}})$. 
This requirement makes sure that the forecaster has no incentive to deviate from his true beliefs to possibly improve the score and is maintained throughout the rest of the paper.

The necessity of giving a clear directive to the forecaster in form of a statistical functional $T$ and to evaluate these forecasts by a suitable consistent scoring function is stressed e.g.\ by \cite{gneiting2011}. A functional is called elicitable if there exists a scoring function that is strictly consistent for it (see \cite{lambert2008} for details). All functionals that are used in this paper are elicitable. A prominent example of a non-elicitable functional is the mode (see \cite{heinrich2013}). Note that usually (infinitely) many consistent scoring functions exist for the same functional (see e.g.\ \cite{ehm2016} for details).

The most prominent example for a consistent scoring function is certainly the squared error for mean forecasts,
\begin{equation} \label{squared_error}
s(x_t,y_t)=(x_t - y_t)^2.
\end{equation}

A consistent scoring function for quantile forecasts is the check function well-known from quantile regression applied to the forecast error, also called quantile score:
\begin{equation} \label{check_loss}
s(x_t,y_t) = \rho_{\tau} ( y_t - x_t ), \text{ where } \rho_{\tau}(u) = u(\tau - \mathds{1}_{\{u<0\}}) \text{ and } \tau \in (0,1).
\end{equation}
A popular example for a proper scoring rule is the logarithmic score: Given a forecast in the form of a density 
$f_t$, i.e. $x_t = f_t$, the logarithmic score is 
$$s(x_t, y_t) = - \log(x_t(y_t)).$$
 
The accuracy of a forecaster or a forecasting method is measured by the expected score or loss
$$\E[s(X_t,Y_t)].
$$

For the example of the squared error expected loss amounts to the mean squared (forecast) error
$$MSE = \E[(X_t-Y_t)^2].$$
For the quantile and the logarithmic score expected loss is represented by the mean quantile score,
$$MQS = \E[\rho_{\tau} (Y_t-X_t)],$$
and the mean logarithmic score,
$$MLS = \E[- \log(X_t(Y_t))],$$
respectively.

Throughout the paper, I will tacitly assume that the expected loss and the other measures of fundamental properties of forecasts $CAL$ and $RES$ and of the process itself $UNC$, which are defined below in terms of expectations over loss functions or differences of loss functions, exist, i.e.\ that the loss functions and the joint distributions of forecasts and observations are sufficiently well-behaved, and that they are constant over time, i.e.\ that the bivariate time series of forecasts and observations fulfils a suitable form of stationarity to guarantee constancy of the occurring unconditional and conditional functionals $T(F_{Y_t})$ and $T(F_{Y_t|X_t})$ showing up later in these terms.   

\subsection{Entropy and Divergence}

To define measures of (mis)autocalibration and resolution and to write down the Murphy decomposition for general loss functions, a general measure of uncertainty for a random variable $V$ and a distance measure between a real number $w$ (which can be interpreted as a forecast of $V$ for our purposes) and a functional of $V$, $T(F_V)$, using consistent scoring functions and proper scoring rules are needed. The generalized entropy and divergence (see \cite{gneiting2007proper} and references therein) are suitable for this purpose.

For a random variable $V$ with distribution function $F_V$ the generalized entropy is defined as 
$$e(V) = \E [s(T(F_V),V)].\footnote{In the literature on proper scoring rules, for some forecast $w$ of a random variable $V$, the short notation $S(w,V) = \E_V [s(w,V)]$ is often used. As this would get cumbersome with the conditioning variables in the time series setting from the next section on and as I want to emphasize the conditioning explicitly , I refrain from doing so.}$$ It measures the expected score achieved by the ideal forecast $T(F_V)$,\footnote{More details on the definition of ideal forecasts follow in the next subsection.} i.e.\ the minimal achievable expected score. It generalizes the classical entropy from information theory as well as the variance and serves as a measure of the inherent uncertainty in $V$: For a distributional forecast in the form of a density function $f$ and the logarithmic score as proper scoring rule the classical entropy $-\E[\log(f_V(V))]$ arises and for a mean forecast with the squared error as a consistent scoring function the variance $\Var(V)$ arises. 

For a forecast $w$ of $T(F_V)$ the divergence between this forecast and the ideal forecast $T(F_V)$ is defined as\footnote{Note that divergence is not a metric, lacking symmetry and the triangle inequality.}
$$d(w, T(F_V)) = \E [s(w, V) - s(T(F_V), V)] = \E[s(w, V)] - e(V).$$ It is the expected difference in scores achieved by $w$ and the optimal forecast. By the definition of consistent scoring functions or proper scoring rules respectively divergence is nonnegative. For the logarithmic score, the Kullback-Leibler divergence $- \E [ \log(\frac{w(V)}{f_V(V)}) ]$ arises as a special case and for the squared error, the squared bias $(w-\E[V])^2$ comes up.

The expected score of $w$ can trivially be decomposed into entropy and divergence, leading to a generalized bias-variance decomposition for the forecasts $w$ of the random variable $V$, which expresses the expected score as the uncertainty in $V$ plus deviations of $w$ from the ideal forecast: 
\begin{equation} \label{basic_decomposition}
\E[s(w,V)] = e(V) + d(w, T(F_V)).
\end{equation}

\subsection{Autocalibration}
 
Calibration describes the statistical compatibility between forecasts and observations. Different notions of calibration exist in the statistical literature, for recent work classifying them and clarifying their relationship see \cite{gneiting2013} and \cite{tsyplakov2014}.

Ideally calibrated or ideal forecasts are constructed from the true conditional distribution, i.e.\ a forecast $X_t$ is ideal relative to the information set $\mathcal{I}_{t-h}$ if
$$X_t=T(F_{Y_t|\mathcal{I}_{t-h}})$$
holds (see e.g.\ \cite{gneiting2013} and \cite{tsyplakov2014}). 

As it is usually difficult to reconstruct the exact information available to the forecaster, a more practicable approach is to condition on $\sigma(X_t)$, the information contained in the forecasts themselves, instead of on $\I_t$:\footnote{Note that I use the usual shorthand notation $Y_t|X_t$ to denote $Y_t|\sigma(X_t)$}

\begin{definition}[Autocalibration, see \cite{tsyplakov2011} and \cite{gneiting2013}] \label{autocalibration}
Forecasts that are ideal with respect to the information contained in themselves,
$$X_t=T(F_{Y_t|X_t}),$$ 
are called autocalibrated.
\end{definition}

Of course $\I_t$ is generally a richer information set, $\sigma (X_t) \subset \I_{t-h}$, and may contain some relevant information that has been overlooked by the forecaster. Nevertheless, in practice this is very hard to assess. Furthermore, ideal calibration implies autocalibration and thus if we detect deviations from autocalibration, the forecasts are not ideally calibrated. In the meteorological literature, autocalibration is simply known as calibration or reliability and is routinely assessed, especially for binary probability forecasts. As will become clear throughout the paper, autocalibration appears naturally within the context of the sharpness principle conjectured by \cite{gneiting2007sharpness} and in the Murphy decomposition. The notion of calibration from the statistical and meteorological literature is closely related to the notion of forecast optimality from the econometric literature (see e.g.\ \citet[chapter 15]{elliott2016} for an overview). Forecast optimality is essentially equivalent to ideal calibration and the popular Mincer-Zarnowitz regression (see \cite{mincer1969}) well-known and widely used in economic forecast evaluation (implicitly) assesses autocalibration of mean forecasts, $X_t = \E[ Y_t | X_t ]$. Other related, but weaker forms of calibration as marginal and probabilistic calibration exist (see again \cite{gneiting2013} or \cite{tsyplakov2014}). 

A natural measure to assess deviations from autocalibration is what I naturally call the conditional divergence between the forecast $X_t$ and the autocalibrated forecast $T(F_{Y_t|X_t})$: $$d(X_t,T(F_{Y_t|X_t})|X_t) = \E[ s(X_t,Y_t) - s(T(F_{Y_t|X_t}), Y_t)|X_t]=\E[ s(X_t,Y_t)|X_t] - e(Y_t|X_t),$$
where I define
$$e(Y_t|X_t) = \E [ s(T(F_{Y_t|X_t}), Y_t)|X_t ]$$ as the conditional entropy, which is disussed in more detail below.

An overall measure of mis(auto)calibration is the expectation of this conditional divergence:
\begin{definition}[Miscalibration] \label{miscalibration}
The overall measure of miscalibration is defined as
$$CAL = \E[d(X_t,T(F_{Y_t|X_t})|X_t)] = \E[ s(X_t,Y_t) - s(T(F_{Y_t|X_t}), Y_t)].$$
\end{definition}
Special cases of $CAL$ have appeared in the literature, see the overview of previous work on special cases of the Murphy decomposition in the introduction. The same holds true for the measure of resolution $RES$ introduced below. $CAL$ is nonnegative due to consistency of scoring functions or propriety of scoring rules respectively and equals zero for autocalibrated forecasts. Note that the first representation of $CAL$ containing the conditional divergence is more useful in terms of theoretical interpretation, while the second term, where the law of iterated expectations has been applied to get rid of the conditioning is more useful in terms of estimation and will be used later on for that purpose. In the example of the squared error, $CAL$ just measures the expected squared conditional bias
$$\E [ (X_t - \E[Y_t|X_t ] )^2 ].$$

In addition to calculating the overall measure of miscalibration $CAL$, miscalibration can also be assessed in detail over the whole range of forecast values. This is best done graphically (at least for point forecasts) by what is called a reliability diagram or calibration plot in meteorology for the special case of binary probability forecasts (see e.g.\ \citet[chapter 7]{wilks2011}), where observed event frequencies are plotted against binned forecasted event probabilities. The generalization to mean or other point forecasts is straightforward by plotting the estimated conditional functionals $T(F_{Y_t|X_t=x})$ against the possible values of the forecasts $X_t=x$. For autocalibrated forecasts the graph is on the diagonal as $x = \E[Y_t|X_t=x]$ holds and for miscalibrated forecasts the miscalibration pattern is illustrated by the deviations from the diagonal.

Assessing autocalibration over the range of values of $X_t$ may uncover systematic mistakes in the forecasts, which could make a correction of these mistakes possible in the future. This process is called recalibration in meteorology. If for example the inflation outcomes averaged to three percent whenever a forecaster forecasted two percent for the mean, this would mean to just always add one percent to his forecasts in these cases to recalibrate him or her. As easy as recalibration sounds in theory, in practice this would require exactly knowing the systematic mistakes, e.g.\ the estimation error and deviations from the true model. Of course this knowledge about the joint distribution of forecasts and observations could be observed over time, but this is probably done by very few forecasters and additional real-world problems like structural change may complicate matters. As will be illustrated in the empirical part of the paper, especially for other distributional features than the mean, it may be difficult to construct autocalibrated forecasts and diverse patterns and large values of miscalibration can be observed in practice.

To assess autocalibration in practice, first the conditional functionals $T(F_{Y_t|X_t})$, e.g.\ $\E[Y_t|X_t]$, have to be estimated. Estimation methods will be discussed in section 4 of this paper.

\begin{example}[Part 1: Introducing the Stylized Forecasters]
As the notions of autocalibration and especially resolution are not widely known in statistics, we use a theoretical example to illustrate the different aspects of forecast quality that they capture. This example builds on an example that has been used by \cite{hamill2001} and \cite{gneiting2007sharpness} in a related context and is expanded here: In a two-step procedure, first the random variable $\mu_t \sim N(0,1)$ is drawn and then $Y_t|\mu_t \sim N(\mu_t, 1)$ realizes independently. This example resembles a situation frequently occurring in practice, where information about the mean of a random variable is known in advance and can be used for forecasting.\footnote{Such a situation could for instance arise if the underlying process was a stationary $AR(1)$ process (without intercept) 
$$Y_t = \phi Y_{t-1} + u_t, \text{ where }|\phi|<1 \text{ and } \{u_t\} \text{ is a white noise sequence},$$
and knowing $\mu_t$ here would mean knowing the process, the value of the parameter $\phi$ and the past of the process.} Consider the following probabilistic forecasters for the random variable at hand,
$$Y_t = \mu_t + \varepsilon_t \text{ with } \mu_t, \varepsilon_t \sim i.i.N(0,1).$$
The first one is a very important reference forecaster, namely the unconditional forecaster, who does not have any knowledge about $\mu_t$ and consequently issues the unconditional distribution of $Y_t$, $N(0,2)$, as his predictive distribution. The second one is the informed forecaster, who does know $\mu_t$ and thus issues $N(\mu_t, 1)$ as forecast distribution. The third one is the sign-reversed forecaster, who issues $N(-\mu_t, 1)$, thus knows $\mu_t$, but makes the systematic mistake of reversing its sign. Observing exact information on the underlying process like the value of $\mu_t$ in our example may not be all that realistic and should rather be an exception than the rule in practice due to measurement error, model misspecification or estimation error. Certainly more realistic is that the forecaster has access to some noisy information. Thus I introduce as the fourth forecaster the noisily informed forecaster as he or she does not observe $\mu_t$, but a noisy version of it, namely $\tilde{\mu}_t = \mu_t + \nu_t$, where $\nu_t \sim N(0,\sigma_\nu^2)$ is independent of $\mu_t$ and $\varepsilon_t$, and issues $N(\tilde{\mu}_t, 1)$ as his predictive distribution. 
Table \ref{table_forecasters} gives an overview of the four different stylized forecasters and their predictive distributions. Two more forecasters, which are represented in the last two rows, will be added to the example soon. 

\begin{table}[]
	\caption{Different probabilistic forecasters $i= Unc, Inf, SR, NI, Rec, Perf$ and the forecast distributions $F_{Y_t}^{pr,i}$ issued by them (Note that I write $N(\mu,\sigma)$ instead of $F_{N(\mu,\sigma)}$ in the table for convenience.)}
	\label{table_forecasters}
	\begin{tabular}{@{}ll@{}}
		\toprule
		Forecaster		& $F_{Y_t}^{(pr,i)} \text{ issued for } Y_t|\mu_t \sim N(\mu_t,1) \text{ with } \mu_t \sim N(0,1)$\\ \midrule
		Unconditional ($Unc$)	& $N(0,2)$ \\
		Informed ($Inf$)	& $N(\mu_t, 1)$ \\
		Sign-Reversed ($SR$)	& $N(-\mu_t, 1)$ \\
		Noisily Informed ($NI$)	& $N(\tilde{\mu}_t, 1) \text{ with } \tilde{\mu}_t=\mu_t + \nu_t, \nu_t \sim N(0,\sigma_\nu^2) \text{ ind.\ of } \mu_t \text{ and } Y_t|\mu_t$\\
		Recalibrated ($Rec$)	& $N(\frac{1}{1+\sigma_\nu^2}\tilde{\mu}_t,  1 + \frac{\sigma_\nu^2}{1 + \sigma_\nu^2})$ \\
		Perfect	($Perf$)& $F^{(pr,Perf)}(x) = \mathds{1}_{\{x \geq Y_t\}}$ \\ \bottomrule
	\end{tabular}
\end{table}

\end{example}

\begin{example}[Part 2: Assessing Autocalibration]

For the popular case of mean forecasts evaluated by the squared error, I calculate the measure of miscalibration $CAL$ for the four forecasters, thus the expected squared conditional bias. The last column of table \ref{table_mean_forecasts} contains these values, while the second column repeats the mean forecasts already contained in table \ref{table_forecasters} for convenience. Details on the calculations can be found in appendix \ref{calculations_example}. The unconditional and the informed forecaster have a miscalibration of 0 as they are autocalibrated and make no systematic mistakes. Whenever they report a forecast $X_t$, the expected outcome given that forecast $\E[Y_t | X_t]$ equals the forecast. Obviously, the unconditional forecasts are not very informative as they do not vary, but this is not captured by calibration. The sign-reversed forecaster makes huge systematic mistakes, which leads to a miscalibration of 4. The noisy mean forecasts $\tilde{\mu_t}$ have a conditional bias of $\frac{\sigma_\nu^2}{1 + \sigma_\nu^2} \tilde{\mu}_t$ for each value of $\tilde{\mu}_t$, leading to a miscalibration of $\frac{\sigma_\nu^4}{1 + \sigma_\nu^2}$. The bias and miscalibration rise with the strength of the noise as expected.

Continuing the example I also present the values of miscalibration $CAL$ for the case of full distributional forecasts in the form of densities evaluated by the logarithmic score in the last column of table \ref{table_distributional_forecasts}. They are qualitatively similar, capturing the same observations as the assessment of the mean forecasts just described. This is as expected since the miscalibration of the sign-reversed forecaster is only caused by the conditionally biased mean forecast and the miscalibration of the noisily informed forecaster is caused by a miscalibrated mean and variance. The other two forecasters show no miscalibration at all. 

The calibration diagram for the example in the case of mean forecasts is plotted in figure \ref{fig:calibrationplot} for a selected value of the noise variance of $\sigma_\nu^2=\frac{1}{2}$. For the calibrated forecasters, i.e.\ the unconditional and the informed forecaster ($i= Unc, Inf$), the graph is on the diagonal and for the noisily informed ($i=NI$) and the sign-reversed ($i=SR$) forecaster the miscalibration pattern is illustrated by the deviations from the diagonal. Of course, in practice the miscalibration pattern can be much more diverse than in this stylized example as will be illustrated in the empirical part of the paper. 

In this stylized example it is easy to correct the systematic mistakes of the miscalibrated forecasters by recalibration, e.g.\ by simply correcting for the conditional bias in the example of the mean forecasts. The noisily informed forecaster could be recalibrated by changing the mean forecast to $\E[Y_t|\tilde{\mu}_t] = \frac{1}{1+\sigma_\nu^2} \tilde{\mu}_t$, which can be interpreted as a linear combination between the unconditional mean (which is 0 in this case) and the noisy conditional mean and puts less weight on the latter the noisier it is. As the noisily informed forecaster also has a miscalibrated variance of 1 not accounting for the noise, the variance also changes for what I name the recalibrated forecaster, who thus issues the distributional forecast $N(\frac{1}{1+\sigma_\nu^2}\tilde{\mu}_t,  1 + \frac{\sigma_\nu^2}{1 + \sigma_\nu^2})$.

\begin{figure}
	\centering
	\includegraphics[width=0.7\linewidth]{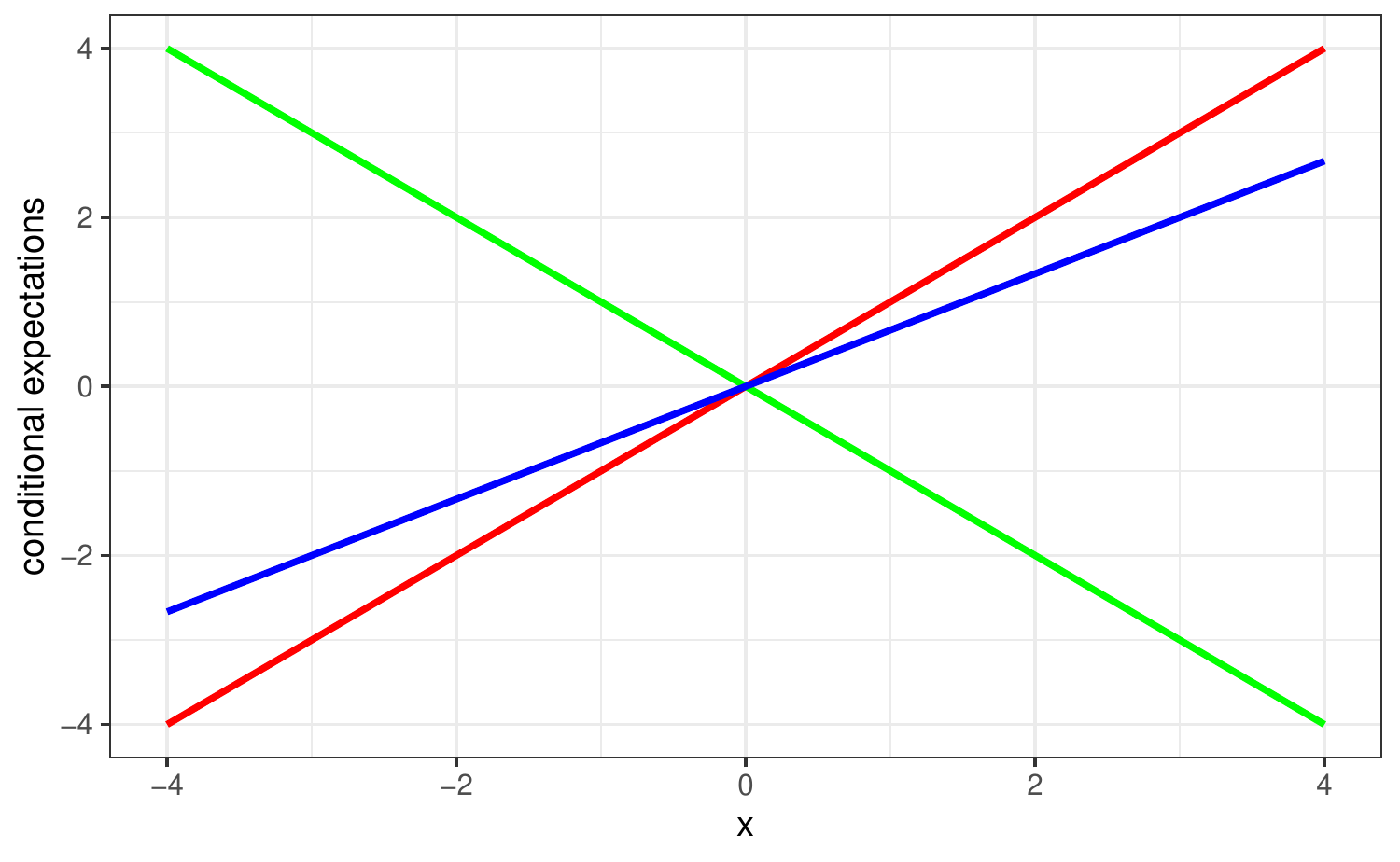}
	\caption{A calibration plot for the forecasters from the example: The lines represent $\E[Y_t|X_t^{(i)}=x]$ for $i= Unc, Inf, SR, NI$, where all the autocalibrated forecasters ($Unc, Inf$) are represented by the diagonal in red (to be precise, $Unc$ is only represented by the origin), the sign-reversed forecaster ($SR$) by the green line and the noisily informed forecaster ($NI$) with noise variance $\sigma^2_{\nu}= \frac 1 2 $ in blue. Miscalibration is indicated by deviations from the diagonal.}
	\label{fig:calibrationplot}
\end{figure}

\end{example}

\subsection{Resolution}

While calibration assesses if the information contained in the forecasts is ideally used or if the forecasts contain systematic mistakes, it does not evaluate if this information is valuable and leads to useful forecasts. Resolution in turn measures exactly this, the ability of forecasts to discriminate between outcome values or, in other words, the information content of the forecasts. The unconditional forecaster from the example is perfectly autocalibrated, but his forecasts are constant and thus not able to discriminate between subsequent outcome values, i.e.\ not informative. The sign-reversed and the noisily informed forecaster in turn are not autocalibrated, but their forecasts contain some valuable information about the outcome, this is captured by resolution. It can be defined as the expected conditional divergence between the unconditional forecast and the autocalibrated forecast:
\begin{definition} \label{resolution}
Resolution is defined as
$$RES = \E[ d(T(F_{Y_t}), T(F_{Y_t|X_t})|X_t) ] = \E[ s(T(F_{Y_t}),Y_t) - s(T(F_{Y_t|X_t}), Y_t)].$$
\end{definition}
As above, the conditional divergence involved is defined as follows: 
\begin{align*} d(T(F_{Y_t}), T(F_{Y_t|X_t})|X_t) &= \E[ s(T(F_{Y_t}),Y_t) - s(T(F_{Y_t|X_t}), Y_t)|X_t]\\
 &=\E[ s(T(F_{Y_t}),Y_t)|X_t] - e(Y_t|X_t).
\end{align*}
Note that again, as for $CAL$, the first representation of $RES$ in definition \ref{resolution} will be used in the theoretical part due to its good interpretability and the second term for the estimation later on. 
Resolution measures how much closer the autocalibrated forecasts $T(F_{Y_t|X_t})$, i.e.\ the ideal forecasts given the information contained in $X_t$, are to the outcomes than the unconditional forecasts not using this information. It thus measures the useful additional variability in the functional of the distribution function of the outcome variable that can be generated by conditioning on the forecasts or the amount of information contained in them. Due to consistency of scoring functions or propriety of scoring rules respectively $RES$ is nonnegative as well. A value of zero means that there is no valuable information contained in the forecasts as the unconditional forecasts not using this information are as good. 

To justify the use of resolution and its interpretation as a measure of information content, I provide two theoretical results: The first result ensures that resolution really does what it is supposed to do, i.e.\ that it is higher for a forecast containing more information.

\begin{proposition} \label{proposition_infocontent}
For two forecasts $X_t^{(1)}$ and $X_t^{(2)}$ with $\sigma (X_t^{(2)}) \subset \sigma (X_t^{(1)})$ it holds that $RES(X_t^{(1)}) \geq RES(X_t^{(2)})$, with equality only if $T(F_{Y_t|X_t^{(1)}})=T(F_{Y_t|X_t^{(2)}})$ almost surely.
\end{proposition}
\begin{proof} See appendix \ref{proofs}. \renewcommand{\qedsymbol}{}	
\end{proof}

The second result is a generalization of the total variance formula
$$ \Var(Y_t) = \Var ( \E [ Y_t|X_t ] ) + \E [ \Var ( Y_t|X_t) ].$$ 
For the case of the squared error loss, resolution equals 
$$\Var ( \E [ Y_t | X_t  ] ) = \E [ ( \E[Y_t|X_t ] - \E[Y_t] )^2 ],$$
i.e.\ the part of the variance of $Y_t$, which is explained by $X_t$. This interpretation is justified by the total variance formula. 
I derive a suitable generalization of this formula, such that the interpretation of resolution as the part of the uncertainty in $Y_t$, in the general case represented by the entropy $e(Y_t)$, which is explained by the forecasts $X_t$, remains valid:
\begin{proposition}[Total Entropy Formula] \label{totalentropy}
The total entropy formula holds:
$$e(Y_t) = \E[ d(T(F_{Y_t}), T(F_{Y_t|X_t})|X_t) ] + \E[ \E [ s(T(F_{Y_t|X_t}), Y_t)|X_t ] ] = RES + \E [e(Y_t|X_t)].$$
\end{proposition}
\begin{proof} See appendix \ref{proofs}. \renewcommand{\qedsymbol}{}	
\end{proof}

The total entropy formula shows that the entropy of $Y_t$ decomposes into resolution and a further term, $\E[e(Y_t|X_t)]$. The term $e(Y_t|X_t)$, which has already been defined above as conditional entropy, and which, following the naming for a special case of it from information theory, can also fittingly be called equivocation of $Y_t$ about $X_t$, measures the uncertainty about $Y_t$ that remains given the knowledge of $X_t$. For the squared error, the equivocation is equal to the second part of the conditional variance formula, the unexplained part of the variance $ \Var(Y_t | X_t ) = \E [ (Y_t - \E [Y_t | X_t] )^2 | X_t ]$, the variance of $Y_t$ that cannot be explained by $X_t$ or that remains given $X_t$. Thus, as in the famous special case, the total entropy in $Y_t$ can be decomposed in a part that can be explained by $X_t$, the resolution $RES$, and in a part that remains unexplained, the expected conditional entropy $\E [e(Y_t|X_t)]$.

As the information content of forecasts has barely been touched upon in statistics and econometrics so far,\footnote{Even in meteorology, where resolution is often used for forecast evaluation via the Murphy decomposition, I am neither aware of an in-depth discussion of its merits as a fundamental property of good forecasts on its own nor of theoretical justifications as I provide here.} with noteworthy exceptions being \cite{holzmann2014}, the two aforementioned papers for the binary case (\cite{galbraith2012}, \cite{lahiri2013}) and the literature on content or predictability horizons (see \cite{galbraith2007}, \cite{isiklar2007} or \cite{diebold2001} and \cite{knueppel2018} for more recent work), I will sketch some of the possible benefits of analyzing information content and especially resolution for forecast evaluation already here before they are illustrated using real-world examples in the empirical part of the paper. 

A first and obvious use of resolution lies in comparing the information content of different forecasts or forecasting methods, for example univariate against multivariate model-based forecasts or model-based against survey forecasts.

A second use that rather judges the absolute usefulness of forecasts is analyzing resolution over the forecasting horizon $h$, where the information content usually decreases with $h$ and often reaches zero rather quickly, making them as useful in terms of information content as the unconditional forecasts $T(F_{Y_t})$. Here it is instructive to normalize the resolution by the entropy, $\frac{RES}{e(Y_t)}$, which leads to a measure of information content, predictability or of variation explained lying between zero and one analogous to the $R^2$ from least-squares regression.\footnote{ Consider again the example of an AR(1) process $Y_t = \phi Y_{t-1} + u_t$. For the optimal mean forecasts $X_t = \E[Y_t|Y_{t-h}]$, this measure equals $\frac{RES}{\Var(Y_t)} = \phi^{2h}$.} In practice a plot of this measure against the forecast horizon $h$ is useful.

It is natural to compare forecasts against an unconditional benchmark (as resolution does) to assess information content or predictability. Actually, \citet[chapter 2]{clements1998} define a random variable $Y_t$ as unpredictable with respect to an information set $\mathcal{J}_t$ if the conditional equals the unconditional distribution, $F_{Y_t|\mathcal{J}_t}= F_{Y_t}$. The definition of resolution is in line with this definition.

The literature on content or predictability horizons mentioned above tries to assess information content as well and strives to find the maximum horizon up to which forecasts contain information about the variable of interest or up to which it is predictable. The used measure there is the quotient between the loss of short-term forecasts and unconditional forecasts (or long-term forecasts, which converge to unconditional forecasts). This is a similar approach to the one sketched above, but, as will become clear in the next section, the effects of miscalibration may disguise these results and they should be reassessed in terms of the more exact measure that we propose, namely (normalized) resolution. As this literature concentrates mainly on mean forecasts, it is certainly a path for future research and very easy by our approach to check more generally for predictability with respect to other features of the distribution. For example, there could be predictability in outer quantiles of stock returns in contrast to the unpredictability at the median or mean.

After having found out that a forecasting method has a low resolution or resolution quickly drops over the forecasting horizon, it is natural to look for the reasons for that, which are certainly hard to analyze empirically. Theoretically, one may categorize these reasons as follows: Are the forecasting methods or the forecasters bad (do they not use the information available to them properly), is their information limited or is the variable not predictable at all? In terms of the our theoretical framework, this may be cast as follows: A low resolution of forecasts may arise due to a bad use of the given information, which would mean that the forecaster did not consider valuable information contained in $\I_{t-h} \setminus \sigma(X_t)$. It may also be that he or she made proper use of the given information $\I_{t-h}$, but lacks information, i.e.\ some valuable information contained in $\mathcal{F}_{t-h} \setminus \I_{t-h}$ is not available to him or her. Finally, it may be the case that he or she has access to all or almost all of the information contained in $\mathcal{F}_{t-h}$, but the information is not useful in predicting $Y_t$, making $Y_t$ unpredictable. Certainly, this is an interesting direction for future research, but requires a detailed knowledge of the forecasting method under analysis and the respective information structure.

\begin{example}[Part 3: Resolution]
	The values for the resolution of the different stylized forecasters can be found in the second to last columns of table \ref{table_mean_forecasts} for the mean forecasts and squared error loss and of table \ref{table_distributional_forecasts} for the density forecasts and logarithmic loss. Details on the calculations can again be found in appendix \ref{calculations_example}. In both cases the unconditional forecaster has a resolution of zero as his constant forecasts are not able to discriminate between the outcomes at all. The informed forecaster has the maximum attainable resolution of 1 or $\frac 1 2 \log(2)$ respectively as he perfectly knows $\mu_t$, the outcome of the first step of the example, and the outcome of the second step, $\varepsilon_t$, is unpredictable. Thus, the informed forecaster can be seen as a reference case that cannot be improved upon in terms of resolution as he possesses the maximum available information. The same holds true for the sign-reversed forecaster, whose forecasts include the same information and who thus has the same resolution, but of course does not use it efficiently and is hence heavily miscalibrated at the same time. 
	
	However, in reality it is difficult or often impossible to assess what the maximum available information is and what part of the variation in the outcome is unpredictable as a not even remotely comprehensible amount of variables may influence the outcome. Thus, such a reference will usually not be available in practice. A reference forecaster that is always available, even though imaginary and not realistic, is the omniscient forecaster, who has the gift of foresight, hence knows the outcome $Y_t$ in advance (even the unpredictable part of the outcome) and issues a degenerate distribution as his probabilistic forecast, i.e.\ $F_{Y_t}^{pr}(x)=\mathds{1}_{\{x \geq Y_t\}}$, and consequently $X_t=Y_t$ as a forecast for the expected value. The resolution of the omniscient forecaster is equal to the entropy of $Y_t$, $e(Y_t)$, which represents a usually unattainable upper bound for resolution. In the case of the squared error the resolution of the omniscient forecaster is thus equal to $\Var(Y_t) = 2$ as he knows not only the outcome of $\mu_t$, but also of $\varepsilon_t$. 
	
	The noisily informed forecaster as well as his or her recalibrated version have a resolution of $\frac{1}{1 + \sigma_\nu^2}$ and $\frac 1 2 \left( \log (2) - \log \left( 1 + \frac{\sigma^2_{\nu}}{1 + \sigma^2_{\nu}} \right) \right) $ respectively, which falls with the amount of noise, always lying between the resolutions of the informed and the unconditional forecaster, which are the limiting cases for $\sigma_\nu^2$ moving towards zero or infinity. 
\end{example}

\section{The Murphy Decomposition, the Calibration-Resolution and the Sharpness Principle}

In the last section I discussed desirable properties of forecasts, focusing mainly on accuracy, autocalibration and resolution. The question naturally comes up if different approaches to forecast evaluation arise from using different properties, which in turn may yield to different results and recommendations. This section answers this question and shows that assessing forecast accuracy and jointly assessing autocalibration and resolution are two sides of the same coin complementing each other perfectly instead of being mutually exclusive approaches to forecast evaluation. I first introduce the generalized Murphy decomposition, then deduce the calibration-resolution principle from it and touch upon some of its implications: Amongst others I point out that it generalizes the sharpness principle and then assess some widely used forecast evaluation methods from its perspective.

\subsection{The Murphy Decomposition and the Calibration-Resolution Principle}

As laid out in the introduction, after having been proposed by \cite{murphy1973} for the special case of probability forecasts for a binary event and having been applied frequently in meteorology as a forecast evaluation method, the Murphy decomposition has been proposed for other types of forecasts and for general classes of loss functions by meteorologists. I provide a fully general version of the decomposition. 

\begin{proposition}[Murphy Decomposition] \label{murphy}
	For a consistent scoring function or a proper scoring rule $s$ it holds that:
	$$ \E[s(X_t,Y_t)] = \underbrace{e(Y_t)}_{UNC} - \underbrace{\E[d(T(F_{Y_t}), T(F_{Y_t|X_t})|X_t)]}_{RES} + \underbrace{\E[d(X_t, T(F_{Y_t|X_t})|X_t)]}_{CAL},$$
	where I call the unconditional entropy uncertainty, $e(Y)=UNC$.
\end{proposition}
\begin{proof} See appendix \ref{proofs}. \renewcommand{\qedsymbol}{}	
\end{proof}

Besides the mechanical and simple version of the proof in the appendix, I sketch a very instructive version here, which involves amongst others a general form of the Sanders decomposition (see \cite{sanders1963}), a predecessor of the Murphy decomposition, which is related in multiple ways to the sharpness principle: I first apply the decomposition from equation \ref{basic_decomposition} of the expected score into an entropy and a divergence term to the random variable $Y_t | X_t$ and the forecasts $X_t$ to decompose the conditional expected score into a conditional entropy term and the divergence between $X_t$ and $T(F_{Y|X_t})$, which measures miscalibration: \begin{align*}
\E[s(X_t,Y_t)|X_t] 
&= e(Y_t|X_t) + d(X_t, T(F_{Y_t|X_t})|X_t).
\end{align*}

Taking unconditional expectations, the second term just yields the measure of miscalibration $CAL$ and I arrive at the fully general version of the decomposition introduced by \cite{sanders1963} for the Brier score:
\begin{equation}\label{sanders}
\E[s(X_t,Y_t)] = \E[ e(Y_t|X_t) ] + CAL.
\end{equation}
The first term, $\E[ e(Y_t|X_t) ]$, was called sharpness by Sanders and is also often referred to as refinement in the meteorological literature (see e.g.\ \cite{jolliffe2012}). Since \cite{gneiting2007sharpness} the term sharpness is used in a more narrow sense as will be explained in the next subsection. \cite{murphy1973} refined this decomposition by decomposing the first term again. This is easily achieved here in the general case by applying the total entropy formula from proposition \ref{totalentropy}.

For the example of the squared error, the Murphy decomposition simplifies and looks as follows:
\begin{equation}\label{murphy_se}
\E[(X_t - Y_t)^2] = \underbrace{\Var(Y_t)}_{UNC} - \underbrace{\Var ( \E [ Y_t|X_t ] )}_{RES} + \underbrace{\E[(X_t - \E[Y_t|X_t])^2]}_{CAL}.
\end{equation}

The Murphy decomposition partitions the expected score of the forecasts $X_t$ into three components: The first one is the unconditional entropy of $Y_t$, which represents the uncertainty in the variable of interest and does not depend on the forecasts. By definition, this term equals the score obtained by the functional of the unconditional distribution of $Y_t$, $T( F_{Y_t})$,
$$e(Y_t) = \E[s(T( F_{Y_t}),Y_t)].$$ 
The second component, resolution, enters the decomposition negatively and represents the part of the uncertainty in $Y_t$ that can be explained by the forecasts, i.e.\ it reduces the expected score by that amount compared to the unconditional forecast. The resulting difference $UNC-RES=\E[e(Y_t|X_t)]$, i.e.\ the expected conditional entropy or expected equivocation of $Y_t$ about $X_t$, represents the minimal achievable score given the information contained in $X_t$, i.e.\ the score achieved by autocalibrated forecasts. For miscalibrated forecasts, the expected score rises by their amount of miscalibration, $CAL$, which is the third component.

What I name the calibration-resolution principle is an immediate consequence of the generalized Murphy decomposition, but has profound consequences for forecast evaluation as I will discuss throughout the remainder of this section.

\begin{corollary}[Calibration-Resolution Principle]
	When constructing forecasts maximizing forecast accuracy, i.e.\ minimizing expected loss $\E[s(X_t,Y_t)]$, is equivalent to jointly minimizing miscalibration $CAL$ and maximizing resolution $RES$. 
\end{corollary}

The calibration-resolution principle as a characterization of forecast accuracy in terms of two underlying properties provides an alternative interpretation and a deeper understanding of what constructing accurate forecasts means and has many implications for the theory and practice of forecast evaluation: It provides a further justification for the accuracy-measurement approach by consistent scoring functions or proper scoring rules put forward by Tilmann Gneiting and his coauthors. Furthermore, it generalizes the sharpness principle conjectured by \cite{gneiting2007sharpness} for probabilistic forecasts to all types of forecasts as will be discussed in detail in the next subsection. The sharpness principle reshaped the field probabilistic forecasting and of probabilistic forecast evaluation by providing a deeper understanding of what a good probabilistic forecast is and which methods are suitable to judge that. The calibration-resolution principle provides such a general perspective for all types of forecasts (including the important case of point forecasts), i.e.\ for forecasting in general. As a consequence it may e.g.\ provide a clearer understanding of the relationships of formerly seemingly unrelated forecast evaluation methods and uncover the drawbacks of several widely used evaluation methods, most often their lack of an assessment of informational content, just as the sharpness principle uncovered the drawbacks of the PIT for the evaluation of probabilistic forecasts. This will be discussed in the subsection after the next. While the calibration-resolution principle makes clear that evaluating forecasts by assessing their accuracy and by analyzing autocalibration and resolution are two sides of the same coin in the sense of leading to results that are in line with each other, it is nevertheless very interesting to flip the coin and to complement average loss by an analysis of autocalibration and resolution: Autocalibration and resolution are not only interesting in their own right as has been laid out in the previous section, but this will also uncover the driving forces of forecasting loss, i.e.\ show if a lack of information content or systematic mistakes or a combination of both are responsible for low accuracy, and where improvements are needed. Thus, the use of the Murphy decomposition as a forecast evaluation method will be the topic of the next section. The example with the stylized forecasters already gives a preview to that.

\begin{example}[Part 4: The Murphy Decomposition]
For the mean forecasts of the stylized forecasters evaluated by squared error loss, the full decomposition is shown in table \ref{table_mean_forecasts}. The uncertainty in the fourth column is of course constant over all forecasters and the resolution and miscalibration in the last two columns have already been discussed in the previous section. They lead to an expected loss or $MSE$ in this case as depicted in the third column. As discussed above, the unconditional forecaster's loss equals the uncertainty. The perfect forecaster has a loss of zero as his resolution equals the uncertainty, while for the more realistic informed forecaster half of the uncertainty is resolved by his forecasts, leading to a $MSE$ of one. The even more realistic noisily informed forecaster has a loss, which is higher by exactly the noise variance $\sigma^2_\nu$ (compared to the informed forecaster) due to a drop in resolution and a rise in miscalibration caused by the noise, while the miscalibration is cured for his recalibrated version, leading to a $MSE$ between the informed and the noisily informed cases. The sign-reversed forecaster, even though having the same resolution as the informed forecaster, has a very high loss due to his massive miscalibration.

\begin{table}[]
	\centering
	\caption{Mean forecasts $X^{(i)}_t=\mu_{Y_t}^{(pr,i)}$ 
		issued by different forecasters $i= Unc, Inf, SR, NI, Rec, Perf$ for $Y_t|\mu_t \sim N(\mu_t,1) \text{ with } \mu_t \sim N(0,1)$ and values of the $MSE$ and the decomposition terms (besides the constant $UNC=2$) for the squared error, $\tilde{\mu}_t=\mu_t + \nu_t, \nu_t \sim N(0,\sigma_\nu^2) \text{ ind. of } \mu_t \text{ and } Y_t|\mu_t$}
	\label{table_mean_forecasts}
	\begin{tabular}{@{}lllll@{}}
		\toprule
		Forecaster & $X^{(i)}_t=\mu_{Y_t}^{(pr,i)}$ & $MSE$ & $RES$ & $CAL$ \\ \midrule
		Unconditional	($Unc$)& 0 &  2     &  0          &  0           \\
		Informed ($Inf$)	& $\mu_t$ &  1             &  1          &   0          \\
		Sign-Reversed ($SR$)	& $-\mu_t$ & 5               &  1          &  4           \\
		Noisily Informed ($NI$)	& $\tilde{\mu}_t$ &   $1+\sigma_\nu^2$     & $\frac{1}{1+\sigma_\nu^2}$        &  $\frac{\sigma_\nu^4}{1+\sigma_\nu^2}$         \\
		Recalibrated ($Rec$)	& $\frac{1}{1+\sigma_\nu^2}\tilde{\mu}_t$ &  1 + $\frac{\sigma_\nu^2}{1 + \sigma_\nu^2} $        &     $\frac{1}{1+\sigma_\nu^2}$       &       0      \\
		Perfect	($Perf$)& $Y_t$ &  0    &   2                  &   0         \\ \bottomrule
	\end{tabular}
\end{table}

\begin{sidewaystable}[]
	\centering
	\caption{Density forecasts $X^{(i)}_t=f_{Y_t}^{(pr,i)}$ issued by different forecasters $i= Unc, Inf, SR, NI, Rec, Perf$ for $Y_t|\mu_t \sim N(\mu_t,1) \text{ with } \mu_t \sim N(0,1)$ and values of the $MLS$ and the decomposition terms (besides the constant $UNC= \frac 1 2 ( \log (4 \pi ) + 1 )$) for the log score, $\tilde{\mu}_t=\mu_t + \nu_t, \nu_t \sim N(0,\sigma_\nu^2) \text{ ind. of } \mu_t \text{ and } Y_t|\mu_t$}
	\label{table_distributional_forecasts}
	\begin{tabular}{@{}lllll@{}}
		\toprule
		Forecaster & $X^{(i)}_t=f_{Y_t}^{(pr,i)}$ & $MLS$  & $RES$ & $CAL$ \\ \midrule
		$Unc$& $f_{N(0,2)}$ &  $\frac 1 2 ( \log (4 \pi ) + 1 )$             &  0          &  0           \\
		$Inf$	& $f_{N(\mu_t, 1)}$ &  $\frac 1 2 ( \log (2 \pi ) + 1 )$              &  $\frac 1 2 \log ( 2 )$     &   0          \\
		$SR$	& $f_{N(-\mu_t, 1)}$ & $\frac 1 2 ( \log (2 \pi ) + 5 )$                &  $\frac 1 2  \log (2)$           &  2           \\
		$NI$	& $f_{N(\tilde{\mu}_t, 1)}$ &  $\frac 1 2 ( \log (2 \pi ) + 1 + \sigma^2_{\nu})$   & $\frac 1 2 \left( \log (2) - \log \left( 1 + \frac{\sigma^2_{\nu}}{1 + \sigma^2_{\nu}} \right) \right) $         &  $\frac 1 2 \left( \sigma^2_{\nu} - \log \left( 1 + \frac{\sigma^2_{\nu}}{1 + \sigma^2_{\nu}} \right) \right) $         \\
		$Rec$	& $f_{N \left(\frac{1}{1+\sigma_\nu^2}\tilde{\mu}_t,  1 + \frac{\sigma_\nu^2}{1 + \sigma_\nu^2} \right)}$ &  $\frac 1 2 \left( \log( 2 \pi ) + \log \left( 1 + \frac{\sigma_\nu^2}{1 + \sigma_\nu^2} \right) \right)$  &           $\frac 1 2 \left( \log (2) + 1 - \log \left( 1 + \frac{\sigma^2_{\nu}}{1 + \sigma^2_{\nu}} \right) \right) $       &       0      \\
		$Perf$& $f_{ N \left( Y_t,  0 \right)}$ &     0                   & $\frac 1 2 ( \log (4 \pi ) + 1 )$  & 0         \\ \bottomrule
	\end{tabular}
\end{sidewaystable}

\begin{figure}
	\centering
	\includegraphics[width=0.7\linewidth]{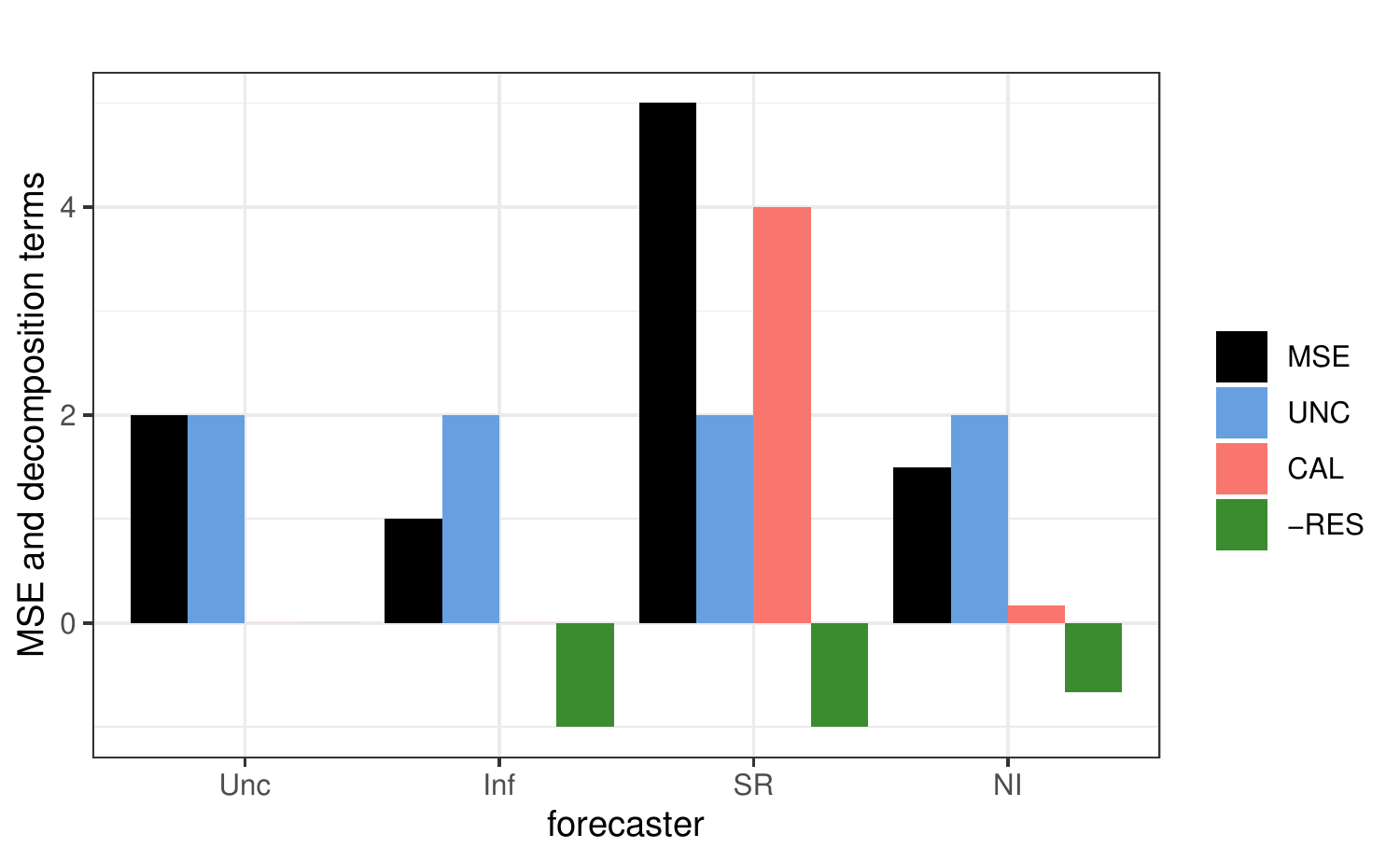}
	\caption{Graphical illustration of the decomposition using squared error loss for the mean forecasts by the unconditional, the informed, the sign-reversed and the noisily informed forecaster, where the noise variance is $\sigma^2_{\nu}=\frac 1 2 $}
	\label{fig:decompositionsexamplesquarederror}
\end{figure}

\begin{figure}
	\centering
	\includegraphics[width=0.7\linewidth]{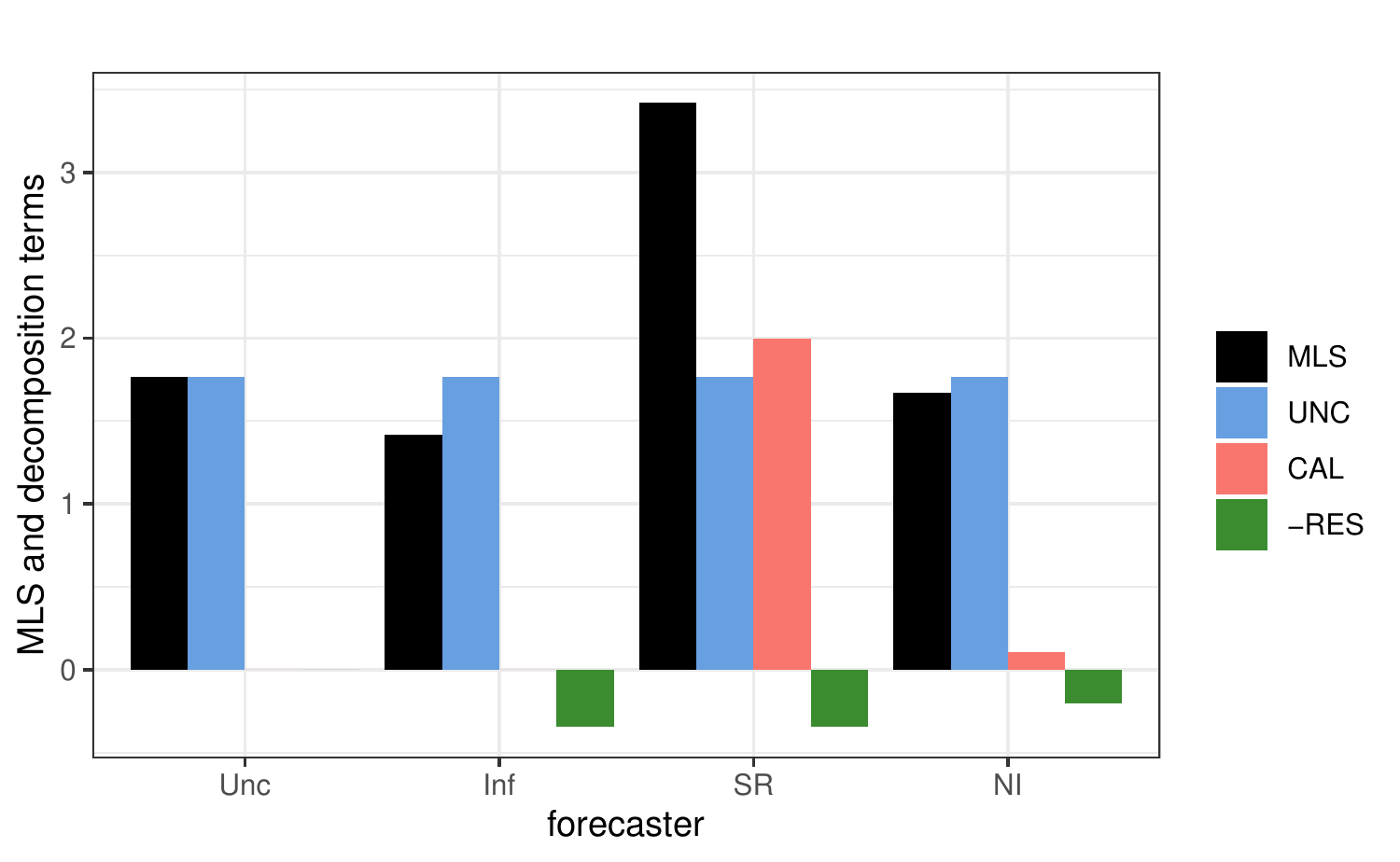}
	\caption{Graphical illustration of the decomposition using the log score for the density forecasts of the unconditional, the informed, the sign-reversed and the noisily informed forecaster, where the noise variance is $\sigma^2_{\nu}=\frac 1 2 $}
	\label{fig:decompositionsexamplelogscore}
\end{figure}

I introduce a graphical representation of the decomposition using bar charts, which facilitates a comparison of its components within and between the single forecasters. Such a plot for the first four forecasters from the example is presented in figure \ref{fig:decompositionsexamplesquarederror} for a noise variance of $\sigma^2_{\nu}=\frac 1 2 $. A similar plot can also and will in the empirical applications be used to compare the development of the decomposition over the forecasting horizon $h$. For the case of density forecasts and logarithmic loss a similar picture emerges as is illustrated in table \ref{table_distributional_forecasts} and figure \ref{fig:decompositionsexamplelogscore}.

\end{example}

\subsection{The Sharpness Principle Revisited}

In this subsection I review the very influential sharpness principle, discuss in which ways the calibration-resolution principle generalizes it and why these generalizations are important for the theory and practice of forecast evaluation.

Sharpness, a property of distributional forecasts popularized by \cite{gneiting2007sharpness} refers to the concentration of the predictive distribution $F_{Y_t}^{pr}$, i.e.\ more concentrated predictive distributions are sharper. Obviously, sharpness alone is not very informative as every degenerate distribution with zero variance is perfectly sharp, but may be far away from the outcomes, i.e.\ strongly miscalibrated.

To measure sharpness, \cite{gneiting2007sharpness} suggest the variance of the forecast distribution, $(\sigma_{Y_t}^{pr})^2$,
as the classical measure of dispersion. Of course other measures of dispersion are also suitable, a natural measure fitting to the specific loss function used is of course the generalized entropy $e((F_{Y_t}^{pr})^{-1}(U_t))$.\footnote{Here, an expression for the random variable having the distribution $F_{Y_t}^{pr}$ is needed to denote its entropy. To obtain it, I use the famous result about the inverse probability integral transform, i.e.\ the fact that for a distribution function $F$ of a continuous random variable it holds that $F^{-1}(U) \sim F$ for $U \sim U[0,1]$. Note that the mean and variance of the predictive distribution, for which the shorthand notations $\mu_{Y_t}^{(pr,i)}$ and $(\sigma_{Y_t}^{(pr,i)})^2$ were used above, could also be expressed in this way as $\E[(F_{Y_t}^{pr})^{-1}(U_t)]$ and $\Var((F_{Y_t}^{pr})^{-1}(U_t))$.}

In their highly influential article \cite{gneiting2007sharpness} conjectured that good or accurate probabilistic forecasts can be characterized as maximizing sharpness subject to calibration.\footnote{To be precise, \cite{gneiting2007sharpness} conjectured the sharpness principle for what they call ideal forecasts, not for accurate forecasts. However, the notion of ideality they use is different to the one used in later work, which I use as well and which is relative to some information set. They define a forecast as ideal if the forecast distribution coincides with the true distribution of the outcome, which can perhaps best be understood as the distribution of $Y_t$ conditional on the largest available information set, which I called $\mathcal{F}_t$. Thus their definition of ideality coincides with ideality with respect to $\mathcal{F}_t$ in our framework. Being ideal with respect to $\mathcal{F}_t$ implies being most accurate and the reverse direction is also true for strictly proper scoring rules. Thus, as the notion of ideality is not entirely precise and if understood in this way equivalent to accuracy anyway, we work with accuracy as \cite{tsyplakov2011} also does.} They discussed several different forms of calibration, which are weaker than autocalibration, and stated the conjecture ``deliberately loosely'' as it was unclear which notion of calibration was needed. This so-called sharpness principle establishes a characterization of accurate forecasts via other properties, namely calibration and sharpness. 
This article has had a profound impact as the paradigm of maximizing sharpness subject to calibration has been widely accepted in the probabilistic forecasting literature and proper scoring rules, the use of which the article also advocates, have become increasingly popular as a forecast evaluation method. 

\cite{tsyplakov2011} was able to prove the sharpness principle after clarifying that autocalibration is the required underlying notion of calibration:\footnote{Note that a prior attempted proof of the sharpness principle based on different notions of calibration (see \cite{pal2009}) turned out to be wrong (see \cite{pal2010}).} Within the general framework used here, the proof is easy using the Sanders decomposition from equation (\ref{sanders}): For a probabilistic forecast in the form of e.g.\ a density or a distribution function, $X_t=T(F_{Y_t}^{pr})$, which is autocalibrated, i.e.\ $F_{Y_t}^{pr} = F_{Y_t|X_t}$, the miscalibration vanishes and the expected score equals the expected conditional entropy. Further, again by autocalibration, the expected conditional entropy equals the expected sharpness (measured by the entropy of the predictive distribution),\footnote{Note that the expected conditional entropy, i.e.\ the measure of sharpness, has the opposite orientation to sharpness. As no confusion should arise I will nevertheless speak of maximizing sharpness when this term is minimized in the following.} i.e.\ in this step the more general notion of sharpness used by \cite{sanders1963} is reduced to the notion of sharpness used by \cite{gneiting2007sharpness}:\footnote{Note that the first step is of course also valid for point forecasts, but the second step is not as it replaces a property of the conditional distribution of $Y_t|X_t$ by a property of the predictive distribution $F_{Y_t}^{pr}$ only. Thus, for point forecasts, this simplification of the expected conditional entropy under autocalibration is not possible. Note however, that for the important special case of mean forecasts and quadratic loss, the interesting relationship $\E[s(X_t, Y_t)] = \Var(Y_t) - \Var(X_t)$ arises.}
\begin{align*} \label{sharpness_principle}
\E[s(X_t, Y_t)] 
&= \E[e(Y_t|X_t)]\\
&= \E[e((F_{Y_t}^{pr})^{-1}(U_t))], \text{ with } U_t \sim U[0,1].
\end{align*}
Thus, for autocalibrated probabilistic forecasts minimizing loss is equivalent to maximizing sharpness. This is a neat characterization of probabilistic forecasts in terms of underlying properties. It also has had a major impact in terms of assessing which evaluation methods are suitable to assess probabilistic forecasts as will be discussed in more detail in the next section.

While \citet[p. 5]{tsyplakov2011} also stressed that ``the sharpness principle provides a useful insight into the essence of probabilistic forecasting'', he nevertheless pointed out that its practical use may be limited as perfectly autocalibrated forecasts are very hard to achieve. I already elaborated further on this and discussed that in practice miscalibration will rather be the rule than the exception as for example information about the underlying process will hardly ever be observed without noise due to for example model uncertainty or estimation error. I introduced the noisily informed forecaster as a stylized example for this. What is more, forecasters may often face a trade-off between using more information and risking miscalibration and not using this information to avoid possible sources of miscalibration. An example for this may be the use of a multivariate model instead of a univariate model, which probably contains more useful information about the underlying process, but is also prone to estimation error and model misspecification. In our stylized example, a forecaster may have to choose between being an unconditional or a noisily informed forecaster, i.e.\ between using the information on $\tilde{\mu}_t$ and not using it. Which choice is optimal will depend on the noise variance $\sigma^2_{\nu}$. This tradeoff cannot be captured by the sharpness principle, which goes along with a maximization of sharpness given perfect autocalibration.    

Hence a generalization of the sharpness principle which is also applicable to miscalibrated forecasts is useful. Furthermore, a generalization to all kinds of forecasts, including point forecasts, and hence stressing the common nature of the forecasting process for point and probabilistic forecasts, is valuable. The calibration-resolution principle yields the desired generalization in both direction, stating that constructing accurate forecasts amounts to jointly minimizing systematic mistakes, i.e.\ mis(auto)calibration, and maximizing informational content, i.e.\ resolution. It replaces sharpness, a property exclusive to probabilistic forecasts, by resolution, a property suitable for all kinds of forecasts, and replaces an optimization conditional on perfect autocalibration by a joint optimization, opening up also to miscalibrated forecasts. This principle is thus an insight into the essence of (all) forecasting and of practical use.

To see that the sharpness principle really is a special case of the calibration-resolution principle, which arises for autocalibrated probabilistic forecasts, remember from above that under autocalibration the expected score, $\E[s(X_t, Y_t)]$, equals the expected conditional entropy, $ \E[e(Y_t|X_t)]$, and that for autocalibrated probabilistic forecasts the conditional entropy equals sharpness, $\E[e((F_{Y_t}^{pr})^{-1}(U_t))], \text{ with } U_t \sim U[0,1]$. Thus, in this case the following relationship holds between resolution and sharpness:
$$RES = UNC - \E[e((F_{Y_t}^{pr})^{-1}(U_t))].$$
Hence, maximizing sharpness is equivalent to maximizing resolution in this case as predictive distributions contain no systematic mistakes under autocalibration and their informational content is higher the more concentrated they are.

From a theoretical point of view, there seems to be no reason to analyze sharpness anymore when evaluating probabilistic forecasts as resolution also captures miscalibrated forecasts, but in practice sharpness is much easier to assess than resolution: Sharpness is a property of the forecast distribution only and can be measured easily, while for resolution a conditional distribution of the observations given the forecasts has to be estimated, which is hard if the forecasts come in form of a full distribution as will be discussed at the beginning of the next section. Thus, sharpness still may play a crucial practical role in the evaluation of probabilistic forecasts. 

\begin{example}[Part 5: Sharpness]
In the example, the perfect forecaster is the sharpest or perfectly sharp as he or she issues a point distribution, but of course infeasible, while the informed forecaster is as sharp as is realistically possible, while still being autocalibrated, as his forecast distribution just represents the unpredictable uncertainty and has just the same variance of 1. The recalibrated forecaster is less sharp, having a variance of $1 + \frac{\sigma_{\nu}^2}{1 + \sigma_{\nu}^2}$. The unconditional forecaster is not very sharp, having a variance of 2. The miscalibrated forecasters both have a variance of 2, but for them sharpness is not a useful measure as discussed above. If we wanted to apply the suitable measure of entropy for the loss function for probabilistic forecasts used before in the example, namely the classical entropy for log loss, this would lead to values of $\frac 1 2 ( \log (2 \pi \sigma^2_{pr,i}) + 1)$, where $\sigma^2_{pr,i}$ stands for the respective predictive variances just discussed, by the formula for the entropy of normally distributed random variable (see appendix \ref{calculations_example}).
\end{example}

%

\subsection{Traditional Forecast Evaluation Methods from the Perspective of the Calibration-Resolution Principle}

The calibration-resolution principle provides a general perspective on forecast evaluation, from which the merits and drawbacks of different forecast evaluation methods can be assessed and the relationships between the methods can be clarified. In order to do this, it is important to clearly show which aspects of forecast quality are captured by certain forecast evaluation methods and which are not. This is facilitated by the use of the three fundamental properties accuracy, resolution and autocalibration and their relationship made clear by the calibration-resolution principle. Using this framework, I point out in this subsection that some widely used forecast evaluation methods do not assess informational content, but only some form of calibration, which is often even weaker than autocalibration. Furthermore, I discuss the relationship between optimality testing and the calibration-resolution principle.

Many traditional and widely used methods of forecast evaluation like the PIT for probabilistic forecasts or conditional coverage or exceedance tests for interval or quantile forecasts focus on assessing some form of calibration. Although undoubtedly being useful tools for forecast evaluation, when used exclusively without accompanying them by the calculation of a suitable loss, these methods yield an incomplete assessment of forecast quality that may leave the forecast examiner clueless or even lead to him or her being content with inferior forecasts or forecasting methods. Such a critique has been issued in recent years for some types of forecasts and the dominant evaluation methodologies used for them. The calibration-resolution principle reveals the common theme here, the lack of an assessment of the informational content, and makes fully transparent what (further) aspects of forecast quality specific methods are missing out on.

The first critique of this form regards the use of the PIT for evaluating probabilistic forecasts. When using the PIT for forecast evaluation (see \cite{dawid1984} and \cite{diebold1998}), the realizations are plugged into the respective forecast distributions and the resulting series of probability integral transforms $F_{Y_t}^{pr}(Y_t)$ is checked for uniformity or uniformity and independence. \cite{hamill2001} and \cite{gneiting2007sharpness} constructed examples 
for forecasts, which are obviously (conditionally) biased, but nevertheless lead to uniform PITs in their simulations. \cite{gneiting2007sharpness} named the uniformity of the PIT probabilistic calibration, stated that this was not enough for forecasts being of a high quality and called for an additional evaluation of sharpness, i.e.\ this critique was the major motivation for them for coming up with the sharpness principle. \cite{mitchell2011} criticized \cite{gneiting2007sharpness} on several ends, questioning amongst others the validity of the sharpness principle and stating that in a proper time series framework and when checking for uniformity and independence (for one-step-ahead forecasts and $h$-independence for $h$-step-ahead forecasts) of the PIT as proposed by \cite{diebold1998}, which they call complete calibration, the relevance of these examples would break down. Even though Mitchell and Wallis make some interesting points, it is clear that even complete calibration is only a necessary condition for ideal calibration. Regarding the question what can go wrong even though forecasts are completely calibrated \cite{tsyplakov2011} points out that this form of calibration is only equivalent to ideal calibration if the information set of the forecaster only consists of the past of the process of interest, i.e.\ if $\mathcal{I}_{t-h}=\sigma(Y_{t-h}, Y_{t-h-1},...)$, and is considerably weaker than autocalibration in realistic settings. Equipped with this knowledge, the calibration-resolution principle reveals that assessing complete calibration is not really an assessment of correct specification or optimality (called ideal calibration here) as intended by \cite{diebold1998}, but only of a weaker form of calibration and ignores the information content of the forecasts. In a related project (\cite{pohle2020probabilistic}) I rename the misleading term complete calibration as autoregressive calibration, come up with realistic examples with an information set $\mathcal{I}_t$ generated by a bivariate process and with empirical examples, where I demonstrate that autoregressively calibrated forecasts may be very bad (i.e. inaccurate) either by a lack of information content or by systematic mistakes (deviations from autocalibration).\footnote{Note that this also shows, where the orignial form of the sharpness principle conjectured by \cite{gneiting2007sharpness} before the clarification of \cite{tsyplakov2011} and the approach to forecast evaluation proposed with it have their weaknesses: An assessment of probabilistic calibration via uniformity of the PIT or the other weak forms of calibration discussed there does not uncover many relevant types of systematic mistakes and, what is more, in this case sharpness is not equivalent to resolution and not really suitable anymore for assessing information content.} This does by no means imply that the PIT is useless as it is still able to uncover systematic mistakes with respect to the information contained in its own past, but that it should not be exclusively used for evaluating probabilistic forecasts as is often done. Instead, it should be accompanied by proper scoring rules and further methods, which come closer to a real check for autocalibration. I discuss new methods going in this direction in \cite{pohle2020probabilistic}.

In a similar fashion as the PIT has been the dominant method for evaluating probabilistic forecasts, the conditional coverage test by \cite{christoffersen1998} has been the dominant method for evaluating interval forecasts. It tests if the sequence of indicator functions, which describe if realizations fell into their respective prediction intervals, is $i.i.Be(p)$, where $p$ is the desired coverage of the prediction interval (again independence should only hold for one-step-ahead forecasts, while for $h$-step-ahead forecasts, $h$-independence should hold). The test is closely related to the PIT and thus checks autoregressive calibration for interval forecasts, thus also ignoring information content and assessing a weaker form of calibration than autocalibration. Hence, using a suitable proper scoring rule, for example the interval score (see e.g.\ \cite{gneiting2007proper}), is advisable here, as the basis for interval forecast evaluation. Complementing this by an assessment of autocalibration and resolution here is possible for interval forecasts as it is for point forecasts. They are only two-dimensional, making estimation of the decomposition terms feasible as will be discussed at the beginning of the next section. The Christoffersen test could then also complement the analysis by these two methods, but should certainly not be used on its own. The test is also often used for quantile forecasts, where the same critique applies.


As a final point in this section, I discuss the relationship between forecast optimality and the calibration-resolution principle. Forecast optimality is a crucial notion mainly in econometrics and there is a huge literature on optimality testing (see e.g.\ \citet[chapter 15]{elliott2016} for an overview) emerging from the rational expectations literature from economics. A forecast $X_t^*$ is defined as optimal with respect to the information set $\I_{t-h}$ and the loss function $l$ if it minimizes expected conditional loss, i.e.\ if
$$  X_t^* = \argmin_{X_t} \E [ l(X_t, Y_t) | \mathcal{I}_{t-h} ] ,$$ 
where $X_t$ are $\mathcal{I}_{t-h}$-measurable random variables. Thus, if $l$ is a consistent scoring function or a proper scoring rule respectively, $X_t^* = T(F_{Y_t|\mathcal{I}_{t-h}})$, i.e.\ the notions of optimality and ideal calibration are equivalent (see also \citet[Theorem 1]{gneiting2011}). If optimality is defined with respect to a smaller information set, this leads to weaker notions of optimality analogous to the weaker notions of calibration mentioned throughout the paper. Optimality is usually tested by checking orthogonality of forecast errors and functions of several variables from $\mathcal{I}_{t-h}$. Well-known problems arising here are that firstly the information set of the forecaster $\mathcal{I}_{t-h}$ is usually unknown and secondly too big, so that it is not entirely clear against which variables to check orthogonality and after orthogonality was not rejected for a set of variables it is never clear if there are other variables which violate it. Thus, here it may be the case as well that only a weaker form of calibration than ideal calibration is tested for. Often optimality is tested against functions of the forecasts $X_t$, which amounts to a test for autocalibration as e.g.\ in the case of the popular Mincer-Zarnowitz regression (see e.g.\ \cite{mincer1969}). Then again the information content would not be captured by the procedure. Arguing similarly for a special field of application, \cite{nolde2017} criticize traditional backtesting of risk measures, which amounts to optimality testing of these risk measures and make a case for comparative backtesting, which amounts to predictive accuracy testing and mention, referring to \cite{holzmann2014}, that traditional backtesting ignores the role of the informational content. 

Evaluating the forecasts by resolution and autocalibration is a good complement to optimality testing as these problems do not arise there. First, the information content of the forecasts is assessed and then the optimality of the forecasts against the information contained in themselves, which is always available as $X_t$ is observed. Furthermore, if optimality is rejected by an optimality test, it is not clear if the deviations from optimality are large (i.e.\ really important or economically significant as an econometrician would say) and in which situation they occur, i.e.\ how the miscalibration pattern looks, and how they can be cured. An assessment of autocalibration answers these questions by providing the shape of the miscalibration pattern with respect to $X_t$ by analyzing the deviations of $T(F_{Y_t|X_t})$ from the diagonal as explained in the previous section and also shows the size of deviations from optimality by the miscalibration $CAL$. Note that in principle the miscalibration pattern and a term measuring overall miscalibration can also be estimated for larger information sets (for example adopted to the information set used in the accompanying optimality test) than $\sigma(X_t)$, but then the estimation becomes high-dimensional, which makes it difficult as discussed in the next section.  

\section{The Murphy Decomposition as a Forecast Evaluation Method}

While the Murphy decomposition has been used to derive the calibration-resolution principle, i.e.\ for theoretical considerations, so far in this paper, as outlined in the introduction, it has originally been proposed as a forecast evaluation method and has been used as such over the years in meteorology. In this section, I discuss estimation of resolution and calibration for point forecasts via nonparametric regression and use the estimated Murphy decomposition to analyze mean forecasts from the Survey of Professional Forecasters and quantile forecasts derived from probabilistic forecasts from the Bank of England. The applications on the one hand demonstrate that analyzing resolution, autocalibration and the Murphy decomposition can lead to useful insights regarding forecast quality also outside the realm of meteorology, in this case in economics, where they had formerly only been applied for the simplest special case of probability forecasts of a binary event, and foreshadow their potential as forecast evaluation methods that could be beneficial in many disciplines. On the other hand, they contain many new insights with respect to the analyzed forecasts, especially the BoE's probabilistic predictions. I also discuss the difficulties associated with the estimation of miscalibration and resolution for probabilistic forecasts and with inference on autocalibration and resolution in general.

\subsection{Estimation} \label{estimation}

To estimate the terms of the Murphy decomposition from proposition \ref{murphy}, the alternative representations of resolution and miscalibration, which get rid of the conditional expectations and which were already mentioned in section 2, can be used. Again, for the interpretation the conditional expectations are useful, but in the estimation unconditional expectations are of course easier to work with. Employing the alternative representations the decomposition looks as follows:
\begin{align} 
\E[s(X_t,Y_t)] =&  \underbrace{\E [s(T(F_{Y_t}),Y_t)]}_{UNC} -  \underbrace{\E[ s(T(F_{Y_t}),Y_t) - s(T(F_{Y_t|X_t}), Y_t)]}_{RES}  \\
& +  \underbrace{ \E[ s(X_t,Y_t) - s(T(F_{Y_t|X_t}), y_t)]}_{CAL}. \nonumber
\end{align}

Apart from the conditional functionals, $T(F_{Y_t|X_t})$, the decomposition is straightforward to estimate: For an observed series of forecasts and realizations of size $N$, $(x_t,y_t)$, $t=1,..., N$, the empirical analogue of the previous equation is

\begin{align}
\frac 1 N \sum_{t=1}^N s(x_t,y_t) =&  \underbrace{\frac 1 N \sum_{t=1}^N s(\widehat{T(F_{y_t})},y_t)}_{\widehat{UNC}} -  \underbrace{ \frac 1 N \sum_{t=1}^N [ s(\widehat{T(F_{y_t})},y_t) - s(\widehat{T(F_{Y_t|X_t = x_t})}, y_t)]}_{\widehat{RES}} \nonumber  \\  
&+  \underbrace{ \frac 1 N \sum_{t=1}^N  [ s(x_t,y_t) - s(\widehat{T(F_{Y_t|X_t = x_t})}, y_t)]}_{\widehat{CAL}}. 
\end{align}

In meteorology, estimators $\widehat{T(F_{Y_t|x_t})}$ based on binning are used here, i.e.\ the range of $X_t$ is divided into bins, for which then the functional is estimated by its empirical analogue over all the values of $y_t$, which belong to an $x_t$ falling into this bin (see e.g.\ \citet{wilks2011} or \cite{ehm2017}). To overcome the deficiencies of these estimators, I use local linear kernel regressions (see e.g.\ \citet{li2007}) in the case of point forecasts, on which I focus in the empirical part of the paper.\footnote{An R-package, which includes functions that estimate the Murphy decomposition for quadratic loss or check loss and provide plots like the barplot illustrating the decomposition terms or calibration plots is currently under development by the author.} Note that \cite{galbraith2011} proposed the use of local constant regressions to analyze autocalibration for probability forecasts of a binary event. I choose the local linear estimator due its advantages over the local constant estimator, which may have e.g.\ a possible large bias at the boundary of support, and much more so over binning estimators, which e.g.\ also may show possibly large biases (see e.g.\ also \cite{li2007}). Since the functional form of $T(F_{Y_t|x_t})$ is usually unknown, nonparametric regression techniques are advisable here, but alternatives from this realm other than kernel regression like e.g.\ splines could also be used.

For the respective consistent scoring function $s$, where I use the squared error (see equation (\ref{squared_error})) to estimate the conditional mean and the check loss (see equation (\ref{check_loss})) to estimate conditional quantiles, for a kernel function $K$, for a bandwidth $H$ and for a sample $(x_t,y_t)$, $t=1,..., N$, find
\begin{equation} \label{kernel_regression}
(\widehat{a(x)}, \widehat{b(x)} ) = \argmin_{a,b} \sum_{t=1}^T s(a + b (x_t - x) , y_t) K \left(\frac{x_t - x}{H} \right)
\end{equation} 
and set $\widehat{T(F_{Y_t|X_t=x})}=\widehat{a}$ to arrive at the local linear estimator.
I use the Gaussian kernel $K(u) = \frac{1}{\sqrt{2 \pi}} \exp{- \frac 1 2 u^2}$ and employ cross-validation with the respective scoring function to select the bandwidth $H$.

(Uniform) consistency of the estimator under quadratic loss, i.e.\ for least-squares kernel regressions, is well-established under suitable assumptions on the underlying stochastic process, i.e.\ weak or strong stationarity and a mixing condition (see e.g.\ \cite{fan1996} or \cite{li2007}). For other loss functions, similar results are to be expected, but not many are established in a time series context. For example for local linear quantile regressions I am not aware of such a result, for the i.i.d.\ case see \cite{fan1994}, see also \cite{yu1998} for a further reference on local linear quantile regression.

In this paper, I focus on point forecasts, i.e.\ on a one-dimensional $X_t$. For probabilistic forecasts $X_t$ is high- or infinite-dimensional, making estimation of the conditional functionals  $T(F_{Y_t|X_t})$ difficult due to the curse of dimensionality. For large samples sizes and for e.g.\ histogram-type forecasts or discrete approximations of continuous forecasts, estimation is possible by multivariate kernel regression. In the aforementioned project (\cite{pohle2020probabilistic}) I also discuss the role of the Murphy decomposition in the evaluation of probabilistic forecasts and how the problem described above could be tackled.

For interval forecasts $X_t$ is two-dimensional and thus estimation of the decomposition with for example the interval score (see \citet{gneiting2007proper}) as a loss function is possible by multivariate kernel regression without the dimensionality problem getting serious. As interval forecasts are closely related to quantile forecasts and the interval score is also derived from the quantile score, the following application on quantile forecasts also foreshadows the use of the decomposition and its usefulness for interval forecasts.

Even though the foci of this paper are different ones and I do not discuss inference on the decomposition terms here, I want to shortly mention some difficulties that arise in this context. When looking at the representations of calibration and resolution in terms of unconditional expectations, it becomes clear that tests on the value of resolution $RES$ or miscalibration $CAL$ are tests for the value of a score difference like the classical Diebold-Mariano [DM] predictive accuracy test (see \cite{diebold1995}). However, the classical DM assumption, a simple high-level assumption on the time series properties of the score difference, leading to standard normality of the test statistic, which comes in form of the average score difference divided by its long-run variance, cannot be used here for several reasons: Here no forecasts are compared, but functionals estimated by kernel regressions, where it is not clear which effect this estimation has on the limiting distribution. Furthermore, in the cases $RES=0$ and $CAL=0$ an analogous problem as in the case of comparing nested models with the DM test arises since the two compared scores are identical apart from the estimation error, see e.g.\ \cite{clark2001} or \cite{mccracken2007}. If the estimation error disappears asymptotically, numerator and denominator of the test statistic go to zero and it is not clear if the test statistic has a degenerate distribution, goes to infinity or is bounded. I hope that future research will tackle the important problem of establishing the limiting distributions of $RES$ and $CAL$ under a suitable estimation approach.

\subsection{Mean Forecasts from the Survey of Professional Forecasters}

The most popular type of forecast in most disciplines is certainly the mean forecast with the squared error as the by far most popular consistent scoring function used for its evaluation. Thus, in my first empirical application I analyze a prominent example of this type, namely the consensus forecasts of inflation and GDP growth from the US Survey of Professional Forecasters.

The SPF is a quarterly survey of macroeconomic forecasts conducted by the Federal Reserve Bank of Philadelphia [Philadelphia Fed]. It is one of the most popular and extensively researched sets of forecasts in economics, see the academic bibliography provided by the Philadelphia Fed for a long list of references.\footnote{https://www.philadelphiafed.org/research-and-data/real-time-center/survey-of-professional-forecasters/academic-bibliographyone} I use inflation and GDP growth consensus forecasts of central tendency, which are created by taking the median of all experts' forecasts.\footnote{An alternative would be to use the mean as the consensus forecast, which usually does not make a big difference (see \cite{mboup2018}). The median is the consensus forecast which is usually used as it robustifies against extreme forecasts by single participants of the survey. However, I also analyzed the consensus forecasts arising by averaging over all individual forecasters and the results were as expected very similar.} The forecasts are for the current and the next four quarters, i.e.\ the forecast horizon ranges from $h=0$ to $h=4$, and the professional forecasters send in their forecasts at about the middle of the quarter, for details on the timing and the survey in general see the in-depth documentation on the SPF.\footnote{https://www.philadelphiafed.org/-/media/research-and-data/real-time-center/survey-of-professional-forecasters/spf-documentation.pdf?la=en} The first forecast-observation pairs in the sample I use are from the third quarter of 1991 (thus having been issued between the third quarter of 1990 and the third quarter of 1991) and the last are for the first quarter of 2019.\footnote{Note that the survey already started in 1968, being the oldest survey of that type in the United States, but due to a change in methodology in 1990 and to not include times with massive structural change in the series, I use this sample period.} I use the latest vintage of realizations available in my dataset.\footnote{The data were retrieved from https://www.philadelphiafed.org/research-and-data/real-time-center/survey-of-professional-forecasters/data-files at 5.07.2019.} Inflation is measured here as annualized quarter-over-quarter percent change of the seasonally adjusted quarterly average price index. 

\begin{figure}
	\centering
	\includegraphics[width=0.7\linewidth]{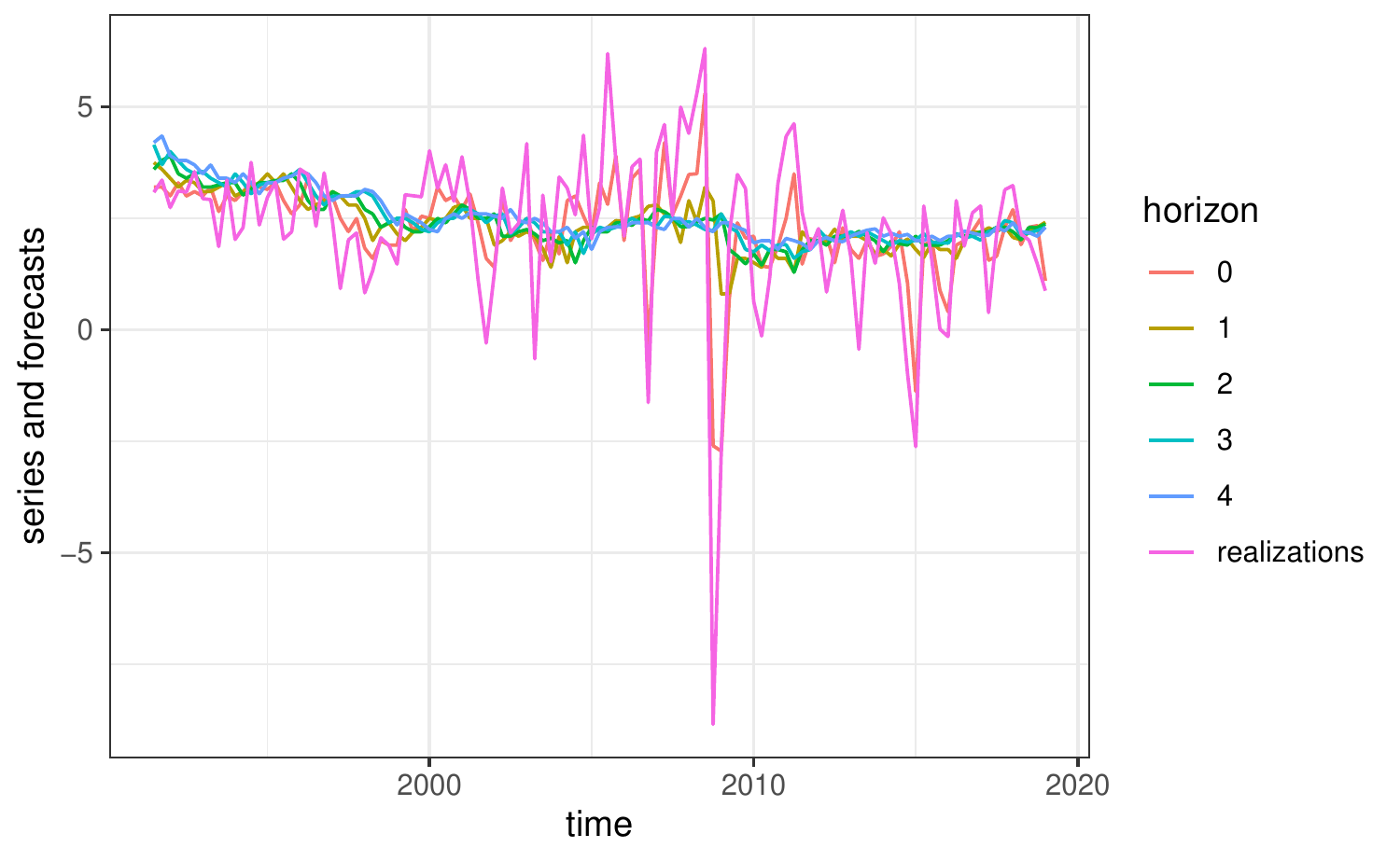}
	\caption{timeplots of the SPF forecasts for horizons 0 to 4 and realizations of CPI inflation from the third quarter of 1991 to the first quarter of 2019 }
	\label{timeplot_SPF_inflation}
\end{figure}

The inflation forecasts for all horizons and the realizations are depicted in figure \ref{timeplot_SPF_inflation}. As expected the forecasts for the current quarter are closer to the outcomes than the forecasts further into the future, which do not show as much variability, which could in turn hint to a lack of resolution. Furthermore, there is a very large drop in inflation due to the financial crisis in the fourth quarter of 2008. However, in a robustness check, where I simply removed the respective period from the sample, the results of interest, i.e.\ the proportions of the decomposition terms, do not change qualitatively, even though the expected loss and the decomposition terms are blown up by the outliers, for details see appendix \ref{robustness_plus_benchmarks}.

As mentioned above, I use the squared error from equation (\ref{squared_error}) as loss function here. Note, however, that the participants of the survey are not explicitly asked for a certain functional of the predictive distribution, thus they might have the mean, the median, the mode or something else in mind. However, it is common practice to assume that these are mean forecasts and to evaluate them by the squared error. As a robustness check, I also did the subsequent analysis with the absolute error $s(x,y) = |x-y|$, which is consistent for the median and which is a special case of the check loss from equation (\ref{check_loss}) used in the next subsection, but the results did not change substantively, see again appendix \ref{robustness_plus_benchmarks} for details. As the Murphy decomposition simplifies to the expression in equation (\ref{murphy_se}) for the squared error, I naturally use the empirical analogue of this simplification for estimation, i.e.\ for a sample of forecasts and realizations of size $N$, $(x_t,y_t)$, $t=1,..., N$,
$$ \frac 1 N \sum_{t=1}^N (x_t - y_t)^2 =  \underbrace{\frac 1 N \sum_{t=1}^N ( y_t - \bar{y})^2}_{\widehat{UNC}} -  \underbrace{ \frac 1 N \sum_{t=1}^N  ( \widehat{\E[ Y_t | X_t ] } - \bar{y})^2}_{\widehat{RES}}  
+  \underbrace{ \frac 1 N \sum_{t=1}^N (\widehat{\E[ Y_t | X_t ] } - x_t)^2 }_{\widehat{CAL}},$$
where $\widehat{\ E[ Y_t | X_t ] }$ is estimated by kernel regression as described above.

\begin{figure}
	\centering
	\includegraphics[width=0.7\linewidth]{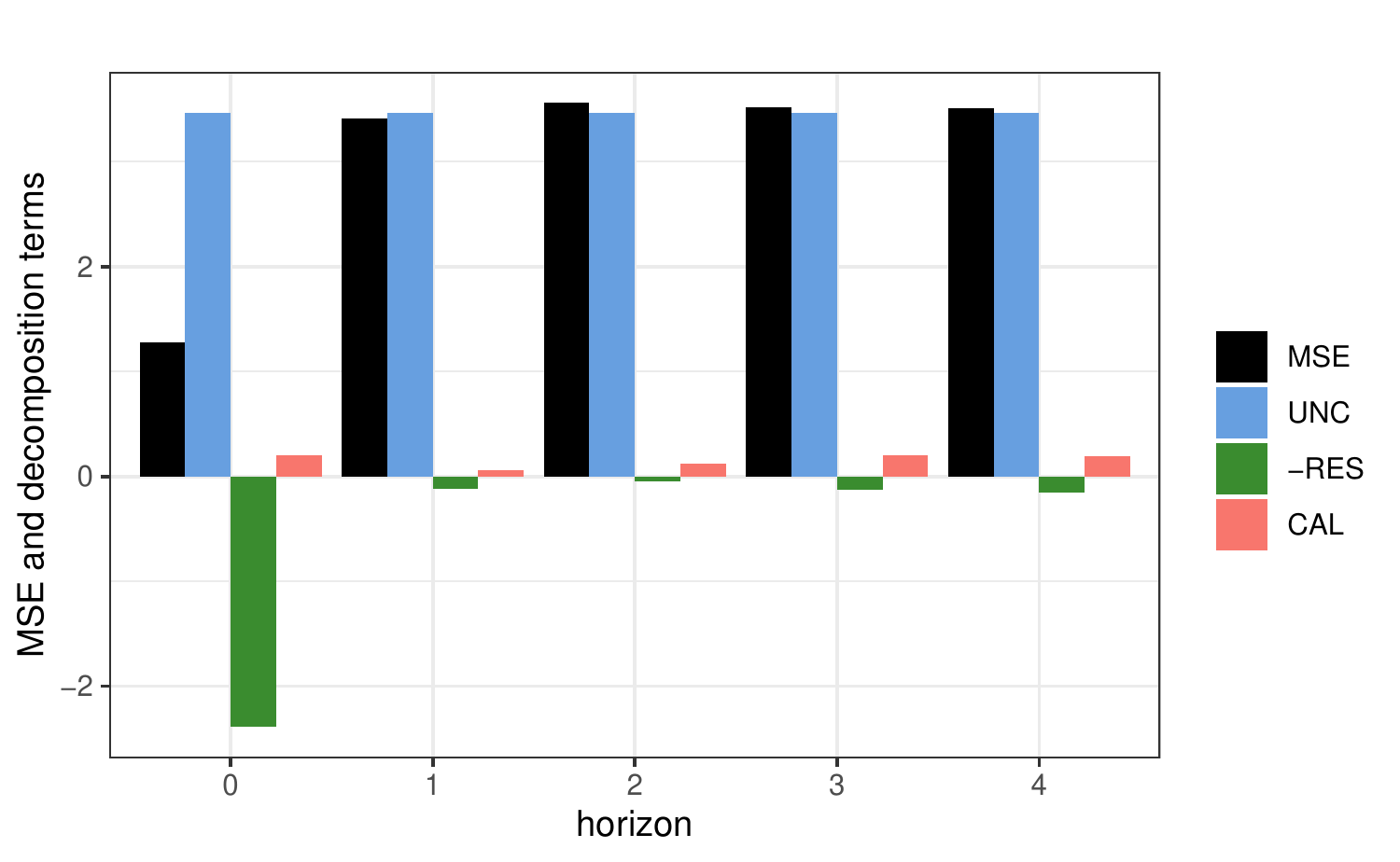}
	\caption{graphical representation of $MSE$ and estimated decomposition terms for quadratic loss for the inflation forecasts of the SPF}
	\label{barplot_decomposition_SPF_inflation}
\end{figure}

The estimated decomposition terms for inflation are reported in table \ref{table_SPF_inflation} and figure \ref{barplot_decomposition_SPF_inflation}. The expected loss, which is the $MSE$ here and which is represented by the black bars in the figure, makes a large jump from the current-quarter forecasts to the one-quarter-ahead forecasts and then stays almost constant for forecasts further into the future. While this behaviour of the $MSE$ of the inflation forecasts is well-known (see for example the documentation of the error statistics by the Philadelphia Fed itself\footnote{https://www.philadelphiafed.org/research-and-data/real-time-center/survey-of-professional-forecasters/data-files/error-statistics}), the estimated Murphy decomposition shows that it is attributable to a drastic drop in information content of the forecasts and not to systematic mistakes made by the forecasters: While the forecasts are able to resolve almost 70 \% of the uncertainty for the current quarter, resolution, represented by the green bars in the figure, virtually drops to zero for a forecast horizon of one quarter and then stays there for forecasts further into the future. Miscalibration, represented by the red bars, is very close to zero for all horizons. 

As explained above, if the expected loss is almost equal to uncertainty $UNC$, here $\Var(Y_t)$, as seems to be the case here from a forecast horizon of $h=1$ on, the forecasts $X_t$ are as good as unconditional forecasts, which arise by using the unconditional mean $\E[Y_t]$. This seems to be a disconcerting result as it means that from the first quarter on forecasts not using any conditioning information are about as good as the SPF forecasts of inflation. However, it has to be kept in mind that the forecasters seem to perform very good in not making any systematic mistakes which would lead to expected losses even higher than uncertainty and forecasts worse than unconditional forecasts. Such forecasts, which do not even meet the requirement of beating the simple unconditional benchmark, arise frequently in practice, especially for long forecast horizons. Rather, the lack of resolution of the forecasts could just be indicative of the difficulty of the forecasting problem at hand. Remember that as discussed in detail in the subsection on resolution in the second section of the paper, there may be three causes for lack of resolution, i.e.\ the forecasters may not be using valuable available information, they may not have enough information or the information may not exist, i.e.\ large parts of the variation in the variable may be unpredictable. As mentioned there, it is hard to empirically decide which factor dominates in a real-world problem. 

However, usually survey forecasts are superior to model based forecasts when it comes to forecasting inflation (see e.g.\ \cite{ang2007}), thus model-based forecasts are not expected to perform better here. At least the classical univariate benchmark in the form of an AR model, which I computed based on real-time data and which the Philadelphia Fed also reports in different specifications as a benchmark, performs much worse in terms of $MSE$ and when the decomposition is applied shows virtually no resolution and a large miscalibration, leading to a $MSE$, which is much larger than uncertainty, for details see as well appendix \ref{robustness_plus_benchmarks}. 

Furthermore, the unconditional benchmark in the form of $\E[Y_t]$ is infeasible and has to be replaced by an estimation with the arithmetic mean of the available past of the series, which usually leads to losses, which are a bit higher than $UNC$, i.e.\ the losses for the infeasible unconditional benchmark. I also report details on this and the results for the feasible unconditional benchmark in appendix \ref{robustness_plus_benchmarks}.  

\begin{table}[]
	\centering
	\caption{$MSE$ and estimated decomposition terms for quadratic loss for the inflation forecasts of the SPF }
	\label{table_SPF_inflation}
	\begin{tabular}{@{}llllllll@{}}
		\toprule
		$h$ &     0      & 1        & 2        & 3  & 4 \\ \midrule
		$MSE$ & 1.28 & 3.41 & 3.56 & 3.52 & 3.51 \\
		$UNC$ & 3.47 & 3.47 & 3.47 & 3.47 & 3.47 \\
		$RES$ & 2.39 & 0.12 & 0.05 & 0.13 & 0.15 \\
		$CAL$ & 0.20 & 0.06 & 0.12 & 0.20 & 0.19 \\
		\bottomrule
	\end{tabular}
\end{table}

\begin{figure}
	\centering
	\includegraphics[width=0.7\linewidth]{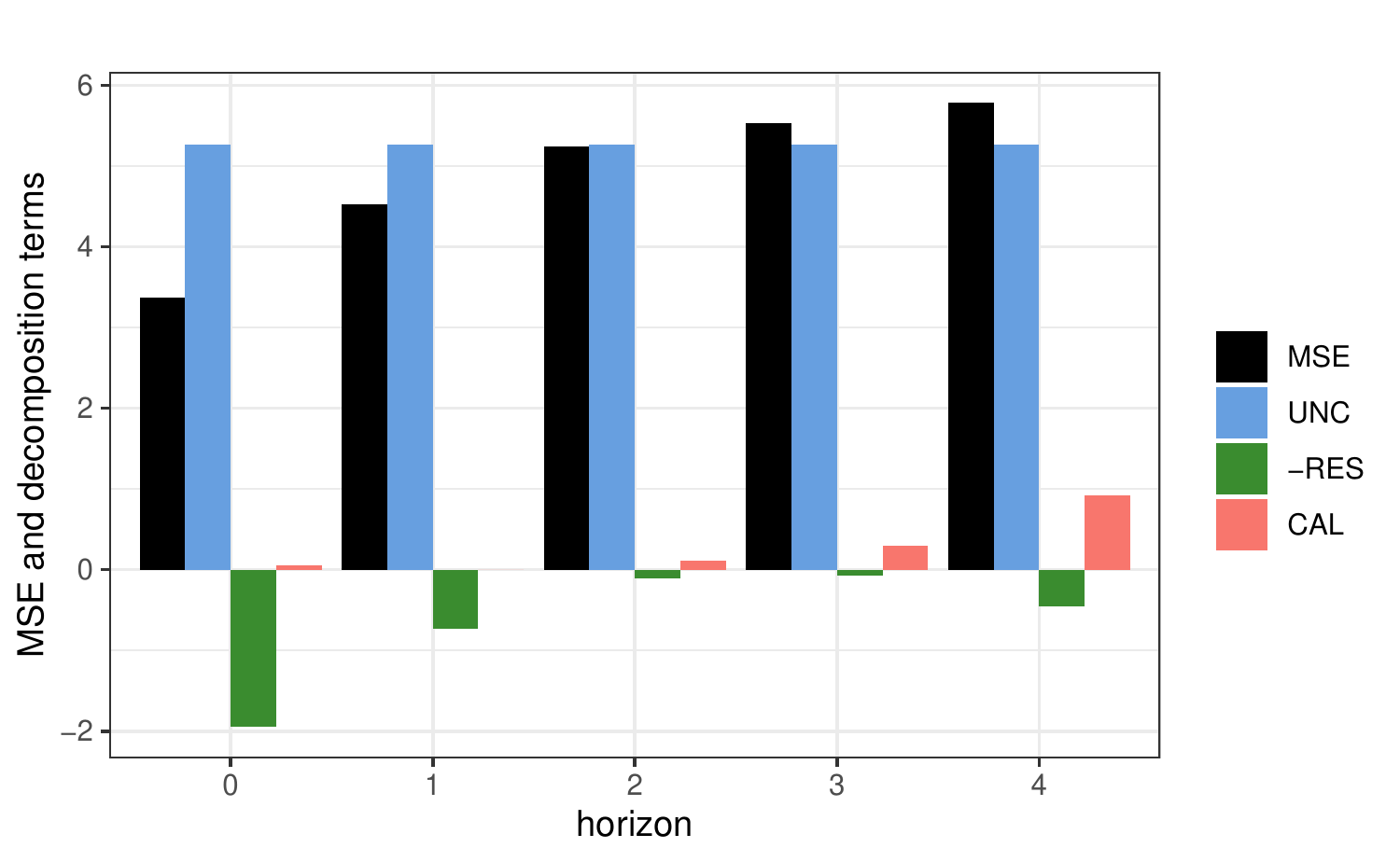}
	\caption{graphical representation of $MSE$ and estimated decomposition terms for quadratic loss for the GDP growth forecasts of the SPF}
	\label{barplot_decomposition_SPF_gdpgrowth}
\end{figure}

For quarterly GDP growth, where also an annualized quarter-on-quarter measure is used, the results are reported in 
figure \ref{barplot_decomposition_SPF_gdpgrowth}. They look very similar to the results for inflation with a slightly more gradual increase in the $MSE$ corresponding to a more gradual drop in resolution.


\subsection{Quantile Forecasts Derived from the Bank of England's Probabilistic Forecasts}

While mean forecasts as treated in the first application are certainly still dominant in most disciplines, other types of forecasts carrying more information on the forecast distribution than only the central tendency have become increasingly popular in recent years. Quantile forecasts are a prime example, appearing for example as a single quantile of interest - e.g.\ as a value at risk in finance, as the bounds of a central prediction interval or in a set of quantiles to characterize the whole predictive distribution. I use one of the most popular examples of probabilistic forecasts in economics, the Bank of England's quarterly forecasts of the distribution of future inflation and GDP growth, calculate quantile forecasts from them and evaluate them by the Murphy decomposition.

The BoE has been issuing probabilistic inflation forecasts since 1996 and probabilistic GDP growth forecasts since 1998.\footnote{In 2016 they also added probabilistic forecasts of the unemployment rate, but here the sample is of course still too short for a proper forecast evaluation.} Since then many central banks have followed this example and have started publishing probabilistic forecasts as well. The forecasts for quarterly inflation and GDP growth come in the form of a two-piece normal distribution, which is governed by three parameters capturing central tendency, dispersion and skewness. For more details on the two-piece normal distribution, its parametrizations, the form in which the BoE issues its forecasts and how to calculate quantiles from this distribution see appendix \ref{split_normal}. The BoE conditions its predictions on assumptions on future interest rates, i.e.\ either on a constant path or on market expectations. I use the latter, but the differences in the predictions are minor and recent work by \cite{knueppel2017} finds that the interest rate assumptions in forecasts of central banks do hardly influence forecast accuracy anyway. As the BoE targeted and forecasted RPIX (Retail Price Index excluding mortgage interest payments) inflation until 2003 and from 2004 on switched to CPI (Consumer Price Index) inflation, I use only CPI inflation. I evaluate forecasts for horizons $h$ of 0 to 6 quarters and the evaluation samples for CPI inflation and GDP growth thus stretch from the third quarter of 2005 and 1999 respectively to the third quarter of 2019, leading to evaluation sample sizes of 57 and 81. The BoE forecasts year-on-year inflation and GDP growth, thus the series are more persistent and should be easier to forecast than the quarter-on-quarter inflation and GDP growth series considered by the SPF. The forecasts are included in the BoE's inflation reports, which have recently been renamed monetary policy reports and which are published in the first days of the second month of a quarter, and the series of historical forecasts are available from the BoE's webpage.\footnote{The whole series of CPI inflation forecasts and the GDP growth forecasts from the fourth quarter of 2007 come with the data for the inflation reports, I downloaded them for the report from August 2019: https://www.bankofengland.co.uk/inflation-report/2019/august-2019. The older forecasts for GDP growth can be downloaded from the national archive https://webarchive.nationalarchives.gov.uk/20170704155503/http://www.bankofengland.co.uk/
publications/Pages/inflationreport/irprobab.aspx; retrieved 5.02.2020.} The respective CPI inflation and GDP series are available from the Office for National Statistics.\footnote{They can be found under the codes D7G7 and ABMI, I use the data published in the fourth quarter of 2019.}

The BoE's probabilistic forecasts have been evaluated several times in the literature, see e.g.\ \cite{wallis2003}, \cite{clements2004}, \cite{mitchell2005}, \cite{gneiting2011comparing}, \cite{galbraith2012} or \cite{straehl2017}. Nevertheless, interesting new and formerly unattainable insights on the quality of those forecasts can be gained employing the Murphy decomposition: I calculate quantiles from the distributional forecasts (see again appendix \ref{split_normal} for details) and then evaluate these quantile forecasts by the check loss from equation (\ref{check_loss}) and the Murphy decomposition, which is estimated as laid out in subsection \ref{estimation}.

\begin{figure}
	\centering
	\includegraphics[width=0.7\linewidth]{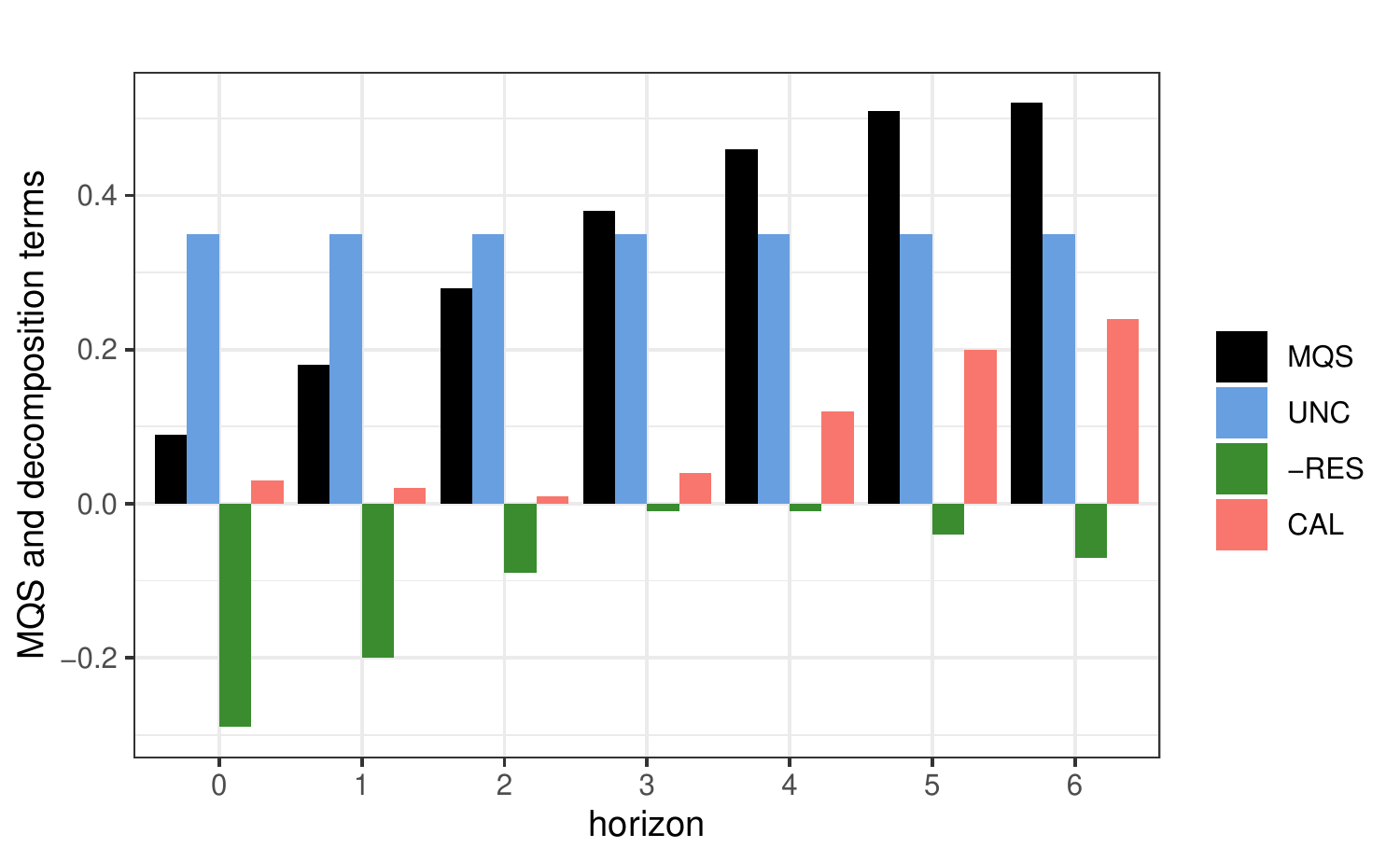}
	\caption{graphical representation of $MQS$ and estimated decomposition terms for the quantile score for the 0.75-quantile forecasts for inflation from the BoE}
	\label{barplot_decomposition_BoE_inflation_tau75}
\end{figure}

In table \ref{table_BoE_inflation_tau75} and figure \ref{barplot_decomposition_BoE_inflation_tau75}, the results for the 0.75-quantile are reported as a first example. The expected loss rises gradually from a very low value relative to the given uncertainty for $h=0$ to a value considerably higher then uncertainty for the longer horizons. This pattern is driven by a decline in resolution from $h=0$ to $h=3$. Initially, it is very high compared to the uncertainty and then declines gradually to about zero for $h=3$ and then stays there. From $h=4$ on the further rise of expected loss is driven by a rise in miscalibration. Thus, the forecasts for the third quartile of inflation are very good up to two quarters into the future, having a high information content and displaying no systematic mistakes, but from four quarters into the future on do not seem to be very useful, lacking information content and displaying considerable amounts of systematic mistakes. 

\begin{table}[]
	\centering
	\caption{$MQS$ and estimated decomposition terms for the quantile score for the 0.75-quantile forecasts for inflation from the BoE}
	\label{table_BoE_inflation_tau75}
	\begin{tabular}{@{}llllllll@{}}
		\toprule
		$h$ &     0      & 1        & 2        & 3  & 4 & 5 & 6 \\ \midrule
		$MQS$ & 0.09 & 0.18 & 0.28 & 0.38 & 0.46 & 0.51 & 0.52 \\
		$UNC$ & 0.35 & 0.35 & 0.35 & 0.35 & 0.35 & 0.35 & 0.35 \\
		$RES$ & 0.29 & 0.20 & 0.09 & 0.01 & 0.01 & 0.04 & 0.07 \\
		$CAL$ & 0.03 & 0.02 & 0.01 & 0.04 & 0.12 & 0.20 & 0.24 \\
		\bottomrule
	\end{tabular}
\end{table}

\begin{figure}
	\centering
	\includegraphics[width=1\linewidth]{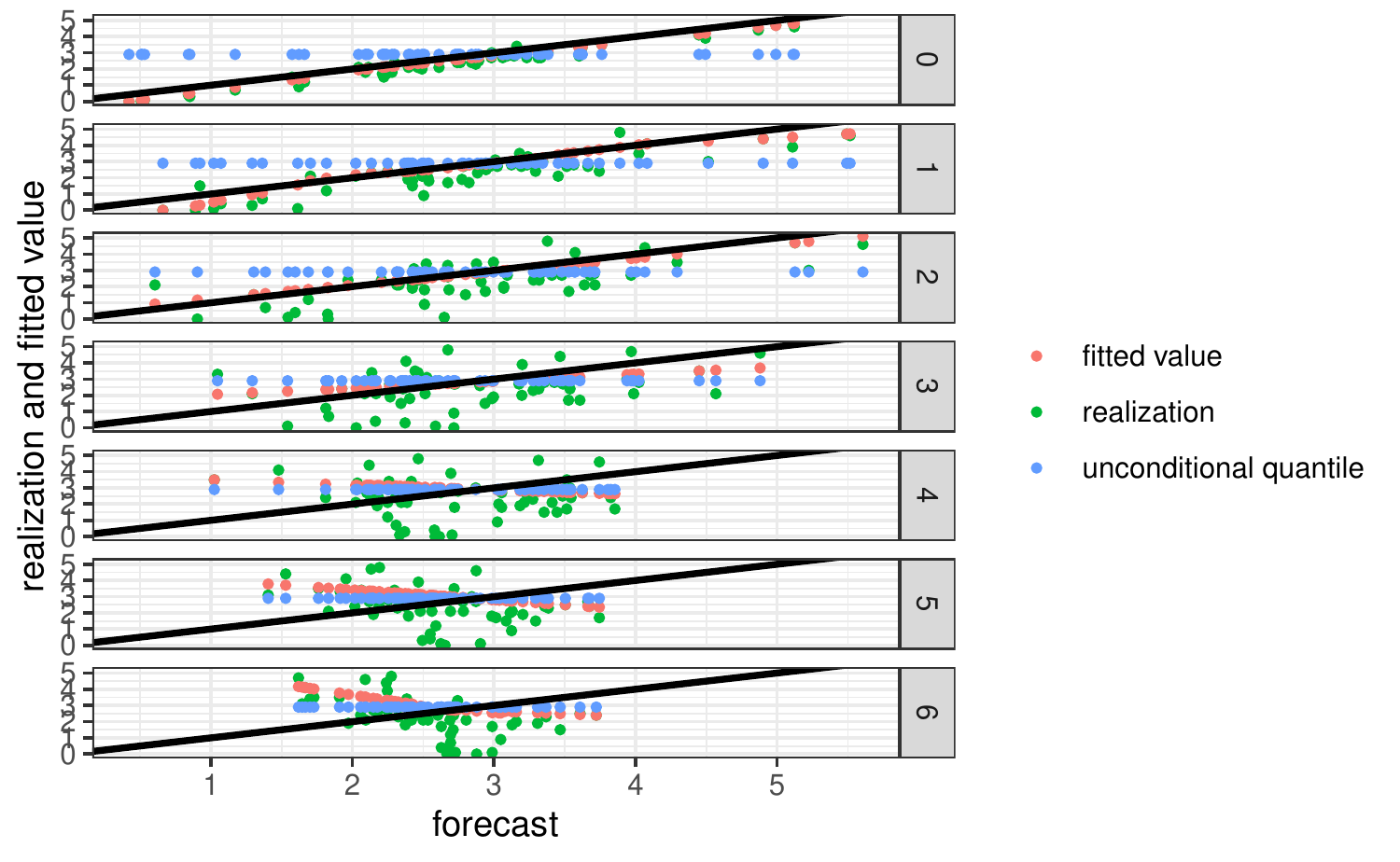}
	\caption{calibration plot for check loss for the 0.75-quantile forecasts for inflation from the BoE}
	\label{calplot_decomposition_BoE_inflation_tau75}
\end{figure}

The pattern of the systematic mistakes is illustrated in detail by the calibration plot in figure \ref{calplot_decomposition_BoE_inflation_tau75}: For the shorter forecast horizons the fitted values, i.e.\ the estimated conditional quantiles obtained by the local linear quantile regression, coloured in red are close to the diagonal leading to a miscalibration close to zero, but for the longer horizons they are close to the horizontal line represented by the unconditional 0.75-quantile coloured blue, even displaying a small negative slope. The negative slope indicates that when higher forecasts for the quantile were issued, this quantile of the subsequent outcomes tended to be lower than in situations when lower forecasts were issued. This miscalibration pattern resembles the sign-reversed forecaster from the stylized example. Furthermore, the fact that the points representing the estimated conditional quantile are close to the estimated unconditional quantile explains the lack of resolution for the longer horizons.  

\begin{figure}
	\centering
	\includegraphics[width=1\linewidth]{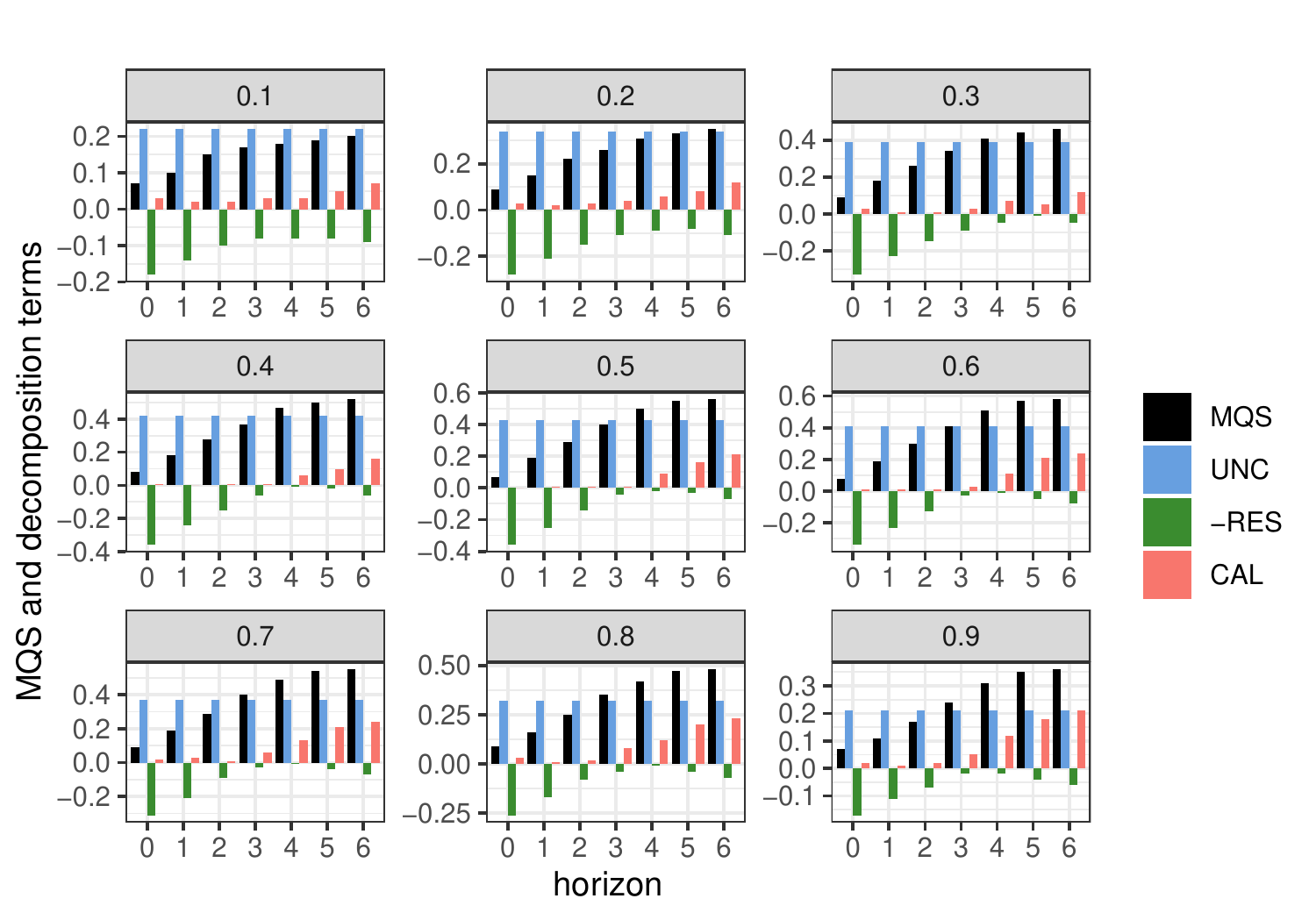}
	\caption{graphical representation of $MQS$ and estimated decomposition terms for the quantile score for forecasts for all deciles for inflation from the BoE}
	\label{barplot_decomposition_BoE_inflation_alldeciles}
\end{figure}

To give an overview of the forecasting performance over the whole forecast distribution in figure \ref{barplot_decomposition_BoE_inflation_alldeciles} the graphical representation of the estimated decomposition is shown for all deciles. As for the third quartile, the expected loss for all deciles is rising gradually with the forecast horizon due to an initially high and then declining resolution and a miscalibration that is very small for the shorter horizons, but rises from about the fourth quarter ahead on. However, there are considerable differences between the deciles in terms of forecast quality: In the lower parts of the predictive distribution the forecasting performance is best and then gradually declines. For the first decile the expected loss is smaller than uncertainty over all considered forecast horizons due to a remarkably high and only slowly declining resolution and only small amounts of miscalibration. With rising deciles resolution declines faster and from about the median on is virtually zero for three or more quarters into the future, but is still substantial for up to two quarters into the future. Further, in the upper half of the distribution considerable miscalibration is observable for four to six quarters into the future, which causes the expected loss to be much higher than the uncertainty, meaning that for the respective quantiles and horizons the unconditional benchmark, here in the form of the respective unconditional quantiles, could do a better job.
Thus, the overall performance of the BoE's CPI inflation forecasts is quite good, showing a high information content for the shorter horizons and for the lower quantiles also for longer horizons. Only at the higher quantiles for the longer horizons there is a substantial amount of miscalibration. 

\begin{figure}
	\centering
	\includegraphics[width=1\linewidth]{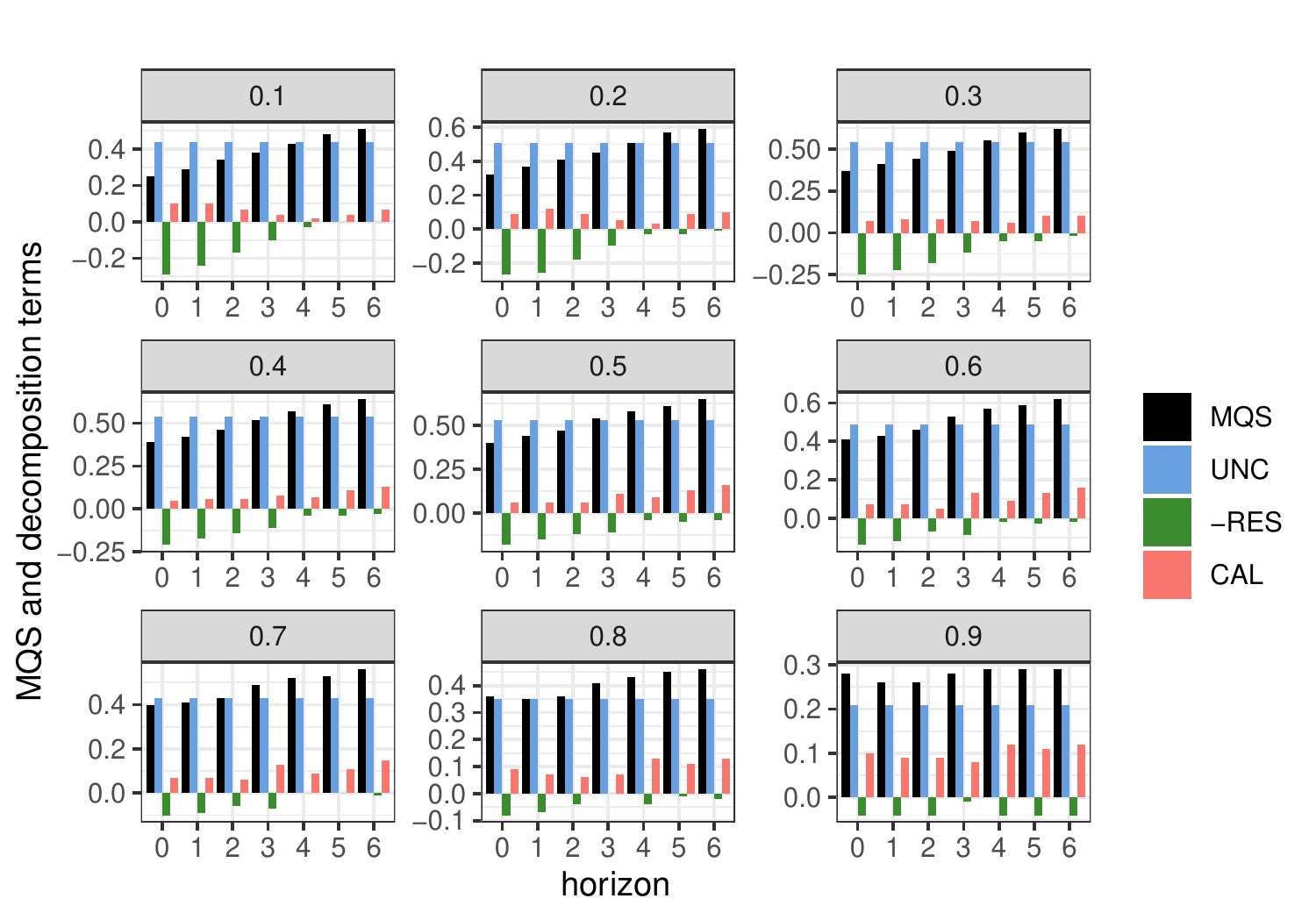}
	
	\caption{graphical representation of $MQS$ and estimated decomposition terms for the quantile score for forecasts for all deciles for GDP growth from the BoE}
	\label{barplot_decomposition_BoE_GDP_growth_alldeciles}
\end{figure}

The results for GDP growth over all deciles are presented in figure \ref{barplot_decomposition_BoE_GDP_growth_alldeciles}. Compared to the inflation forecasts the information content is much smaller overall. A similarity is that the forecasts are best at the lower deciles, i.e.\ show a higher information content, smaller amounts of miscalibration and thus a lower expected loss compared to the uncertainty, and that the quality of the forecasts decreases gradually when moving to the upper deciles: While for the first three deciles there are substantial amounts of resolution up to a forecast horizon of three quarters and only small amounts of miscalibration for all forecast horizons, for the last two deciles there is virtually no resolution accompanied by some miscalibration such that the expected loss is larger than uncertainty for all forecast horizons. Thus, the overall performance of the GDP growth forecasts is not as convincing, especially not in the upper parts of the forecast distribution, while in the lower parts they still contain some useful information. 

These results give new insights regarding the quality of the BoE's probabilistic forecasts of inflation and GPP growth with regards of course to information content and systematic mistakes, but also the analysis of different parts of the forecast distribution. These insights are obtained without even comparing them to competing forecasts as would be required for most other forecast evaluation methods. 

Nevertheless, a comparison to competing forecasts is of course useful and can lead to further insights. Therefore, in recent work (\cite{pohle2020benchmark}) I compare the BoE's probabilistic forecasts to forecasts from two methods, which I propose as natural benchmarks for quantile forecasting: The first is a classical AR model plus a normality assumption on the innovations, which carries over to the normality of the prediction intervals, see e.g.\ \citet[chapter 5]{box}, and the second is the QAR model by \cite{koenker2006}. For inflation the benchmark methods both show a similar performance, but cannot compete with the BoE's forecasts. For GDP growth the AR model is with a huge margin outperformed by the other two, while the QAR model shows a good performance compared to the BoE's forecasts, even dominating it for some horizons and in some parts of the distribution.

%

\section{Conclusion}

The basis of this paper is the fully general version of the Murphy decomposition, which I provide. It expresses the expected loss of forecasts in terms of a component representing the inherent uncertainty of the process of interest, resolution and miscalibration. Resolution measures information content and miscalibration measures systematic mistakes, i.e.\ deviations from autocalibration. The Murphy decomposition originates from meteorology, where it was proposed and is widely used as a forecast evaluation method.

In section 3 I make use of the decomposition in a different way, using it for theoretical insights. I introduce the calibration-resolution principle as a direct consequence of the decomposition. It states that constructing accurate forecasts amounts to jointly maximizing resolution and minimizing miscalibration. This simple and seemingly obvious principle has powerful implications, some of which I then discuss: It generalizes the influential sharpness principle by \cite{gneiting2007sharpness} from autocalibrated probabilistic forecasts to all kinds of forecasts, characterizing accurate forecasts in terms of fundamental underlying properties. From this general perspective on forecast evaluation I assess the role of some widely-used forecast evaluation methods and amongst others point out that many of them do not assess information content at all and should be accompanied by appropriate complementary methods.

Prior to that, in section 2, I carefully define and illustrate the important properties of autocalibration and resolution, which are already of considerable interest in their own right, but have been neglected in the literature. I also discuss their role for forecast evaluation.

In section 4 I use the Murphy decomposition for its original purpose, as a forecast evaluation method. I discuss its estimation, proposing amongst others to use more advanced nonparametric regression techniques, e.g. local linear (in my applications least-squares or quantile) regressions, than the simple binning estimators used in meteorology to estimate the appearing conditional functionals. Then I apply the Murphy decomposition to two sets of popular economic mean and quantile forecasts respectively, which seems to be the first application of the decomposition - with the exception of the case of binary probability forecasts - in economics and probably even outside of meteorology. The evaluation of the SPF mean forecasts of inflation and GDP growth reveals that the current-quarter forecasts have a high information content, which drops virtually to zero for longer horizons, while there are almost no systematic mistakes in the forecasts. Classical benchmarks are clearly outperformed, suggesting no immediate room for improvement. The evaluation of the BoE's probabilistic forecasts of inflation and GDP growth over a range of quantiles brings with it many interesting insights with respect to the quality of these forecasts: While for both variables there is usually considerable information content for shorter horizons and varying amounts of miscalibration for longer horizons and a decrease in forecast quality with rising quantile levels, the inflation forecasts are in general much better in terms of information content and expected loss compared to uncertainty. 
The applications illustrate that the Murphy decomposition may yield valuable additional and formerly unattainable insights with respect to information content, systematic mistakes and other aspects of forecast quality, while being easy to implement, report and understand. Thus, it is a valuable complement to other forecast evaluation methods, which may contribute substantially to the ultimate goal of forecast evaluation, the improvement of future forecasts.

Given the demonstrated potential of the calibration-resolution principle from a theoretical perspective and the Murphy decomposition as a forecast evaluation method from a practical perspective, many interesting directions for future research arise: Inference on the decomposition terms and the estimation of the decomposition for probabilistic forecasts are two of them. Applications of the decomposition to other types of forecasts, e.g.\ expectile forecasts, and to other data sets, including cross-sectional problems and large data sets, are of obvious interest as well. Moreover, as discussed in the subsection on resolution, the maximum forecast horizon, up to which forecasts contain valuable information, and closely related to this the limits of predictability for a certain variable are promising research targets in many disciplines. A further topic of interest is forecast combination, where it is interesting to look at the decomposition to e.g.\ find out if combining the forecasts rather reduces miscalibration through averaging out mistakes or increases resolution by combining different information sources. A related topic is the application of the decomposition to survey forecasts, on the one hand to the single participants, on the other hand to consensus forecasts.

%







\bibliography{literature_decomp}

\appendix
\appendixpage

\section{Proofs of Propositions} \label{proofs}

\begin{proof}[Proposition 1]
I prove that for a strictly consistent scoring function or a strictly proper scoring rule $s$ and two forecasts $X_t^{(1)}$ and $X_t^{(2)}$ with $\sigma (X_t^{(2)}) \subset \sigma (X_t^{(1)})$ it holds that $RES(X_t^{(1)}) \geq RES(X_t^{(2)})$, with equality only if $T(F_{Y_t|X_t^{(1)}})=T(F_{Y_t|X_t^{(2)}})$ almost surely.
\begin{align*}
&\ \ \ \ RES(X_t^{(1)}) - RES(X_t^{(2)})\\
&= \E[ s(T(F_{Y_t}),Y_t) - s(T(F_{Y_t|X_t^{(1)}}), Y_t)] - \E[ s(T(F_{Y_t}),Y_t) - s(T(F_{Y_t|X_t^{(2)}}), Y_t)]\\
&= \E[ s(T(F_{Y_t|X_t^{(2)}}), Y_t) - s(T(F_{Y_t|X_t^{(1)}}), Y_t)]\\
&= \E[\E[ s(T(F_{Y_t|X_t^{(2)}}), Y_t) - s(T(F_{Y_t|X_t^{(1)}}), Y_t)|X_t^{(1)}]] \geq 0,
\end{align*}
where in the last step the definition of strictly consistent scoring functions or strictly proper scoring rules respectively is used and thus equality only holds when\\ $\mathcal{P}(T(F_{Y_t|X_t^{(1)}})=T(F_{Y_t|X_t^{(2)}}))=1$. 
\end{proof}

\begin{proof}[Proposition 2]

To arrive at the total entropy formula note that expected conditional entropy can be expressed as the difference between the unconditional entropy and the negative of $RES$, which can then be rearranged: 
\begin{align*}
\E [e(Y_t|X_t)] &=  \E [ \E [ s(T(F_{Y_t|X_t}), Y_t)|X_t ] ]\\
&=  \E [ \E [ s(T(F_{Y_t}), Y_t)|X_t ] ] - \E [ \E [ s(T(F_{Y_t}), Y_t) - s(T(F_{Y_t|X_t}), Y_t)|X_t ] ]\\ 
&= e(Y_t) - \E[ d(T(F_{Y_t}), T(F_{Y_t|X_t})|X_t) ].
\end{align*}

\end{proof}

\begin{proof}[Proposition 3]

By adding zero twice and rearranging, in the first step I arrive directly at the alternative representation of the decomposition, where the second and third term just represent the alternative representations of $CAL$ and $RES$ from definitions \ref{miscalibration} and \ref{resolution}, which explains the second equality:
\begin{align*}
&\E[s(X_t,Y_t)]\\
&= \E [s(T(F_{Y_t}),Y_t)] - \E[ s(T(F_{Y_t}),Y_t) - s(T(F_{Y_t|X_t}), Y_t)] + \E[ s(X_t,Y_t) - s(T(F_{Y_t|X_t}), Y_t)]\\
&=\underbrace{e(Y_t)}_{UNC} - \underbrace{\E[d(T(F_{Y_t}), T(F_{Y_t|X_t})|X_t)]}_{RES} + \underbrace{\E[d(X_t, T(F_{Y_t|X_t})|X_t)]}_{CAL}.
\end{align*}

\end{proof}

\section{Calculations for the Example} \label{calculations_example}

In this part of the appendix, I provide the calculations underlying the theoretical example used throughout sections 2 and 3 of the main text.
For convenience, we first repeat some fundamental definitions and results from the main text. For a random variable $V$, the entropy is defined as 
$$e(V) = \E [s(T(F_V),V)]$$ and the divergence between a forecast $w$ of the functional $T(F_V)$ and this functional as 
$$d(w, T(F_V)) = \E [s(w, V) - s(T(F_V), V)] = \E[s(w, V)] - e(V).$$
The expected loss can be decomposed into
$$\E[s(X_t,Y_t)] = \underbrace{e(Y_t)}_{UNC} - \underbrace{\E[d(T(F_{Y_t}), T(F_{Y_t|X_t})|X_t)]}_{RES} + \underbrace{\E[d(X_t, T(F_{Y_t|X_t})|X_t)]}_{CAL}.$$

For the squared error $s_{SE} (x_t, y_t) = (y_t - x_t)^2,$ the entropy and divergence are
$$e_{SE}(V) = \Var(V)$$
and
$$d_{SE} (w, \E[V]) = (w - \E[V])^2,$$
leading to the the following resolution, calibration and decomposition:
$$\E[(X_t - Y_t)^2] = \underbrace{\Var(Y_t)}_{UNC} - \underbrace{\Var ( \E [ Y_t|X_t ] )}_{RES} + \underbrace{\E[(X_t - \E[Y_t|X_t])^2]}_{CAL}.$$

For the forecasters from the example described in detail in the main text and summarized in table \ref{table_forecasters}, this leads to the values of the decomposition components for squared error loss (and thus for mean forecasts) as contained in table \ref{table_mean_forecasts}. I present some details on the calculations here: The uncertainty amounts to 
$$UNC=\Var(Y_t) = \Var(\mu_t) + \Var(\varepsilon_t) = 2.$$

For the unconditional forecaster, the values of the decomposition terms are straightforward to calculate noting that $X_t= \E[Y_t | X_t] = \E[Y_t]=0$.
For the informed forecaster, $CAL=0$ as $X_t=\E[Y_t|X_t]=\mu_t$ and $RES=\Var(\mu_t)=1$ and $MSE=\Var(\varepsilon_t)=1$.

For the sign-reversed forecaster $\E[Y_t|X_t] = \mu_t$ as well, thus resolution does not change, $RES=1$, but miscalibration rises to $CAL = \E[-2 \mu_t]=4$, leading to very inaccurate forecasts, $MSE=5$, by the decomposition formula. 

For the noisily informed forecaster, who observes a noisy version of $\mu_t$, namely $\tilde{\mu}_t= \mu_t + \nu_t$, where $\nu_t \sim N(0, \sigma^2_{\nu})$, and issues this as his mean forecast, I calculate $\E[Y_t|X_t]$:
\begin{align*}
\E[Y_t|X_t] &= \E[\mu_t + \varepsilon_t| \tilde{\mu}_t]\\
			&= \E[\mu_t | \tilde{\mu}_t]\\
			&= \frac{\Cov(\mu_t, \tilde{\mu}_t)}{\Var({\tilde{\mu}_t})} \tilde{\mu}_t\\
			&= \frac{\Var(\mu_t)}{\Var({\tilde{\mu}_t})} \tilde{\mu}_t\\
			&= \frac{1}{1 + \sigma^2_{\nu}} \tilde{\mu}_t.			
\end{align*}
Here I used in the third step that if $V$ and $W$ are jointly normal,
$$\begin{pmatrix} V \\ W \end{pmatrix} \sim N \left( \begin{pmatrix} \mu_V \\ \mu_W \end{pmatrix}, \begin{pmatrix} \sigma^2_V & \rho \sigma_V \sigma_W   \\ \rho \sigma_V \sigma_W & \sigma^2_W \end{pmatrix} \right),$$
then 
$$V|W \sim N \left( \mu_V + \frac{\sigma_V}{\sigma_W} \rho (W - \mu_W), (1 - \rho^2) \sigma^2_V \right).$$
Note that $\mu_t$ and $\nu_t$ are independent normal and thus jointly normal random variables and that $\mu_t$ and $\tilde{\mu}_t$ are as well jointly normal as linear combinations of jointly normal random variables are jointly normal.
This leads to a miscalibration of
$$CAL = \E \left[ \left(\tilde{\mu}_t - \frac{1}{1 + \sigma^2_{\nu}} \tilde{\mu}_t \right)^2 \right] = \left(\frac{\sigma_{\nu}^2}{ 1 + \sigma_{\nu}^2} \right)^2 \Var(\tilde{\mu}_t) = \frac{\sigma^4_{\nu}}{ 1 + \sigma_{\nu}^2},$$
a resolution of
$$RES = \Var \left(\frac{1}{1 + \sigma^2_{\nu}} \tilde{\mu}_t \right) = \frac{1}{ (1 + \sigma_{\nu}^2 )^2} \Var(\tilde{\mu}_t) = \frac{1}{1 + \sigma_{\nu}^2}$$
and an expected loss of
$$MSE = \E[((\mu_t + \nu_t) - (\mu_t + \varepsilon_t))^2] = \E[( \nu_t - \varepsilon_t))^2] = 1 + \sigma^2_{\nu}.$$

For the recalibrated version of the noisily informed forecaster the miscalibration of course disappears and the resolution does not change, leading to more accurate forecasts, i.e.\
$$MSE = 2 - \frac{1}{1 + \sigma_{\nu}^2} = 1 + \frac{\sigma^2_{\nu}}{1 + \sigma^2_{\nu}}.$$

For the perfect forecaster, the values are again straightforward to calculate noting that $X_t=\E[Y_t|X_t]=Y_t$.

For density forecasts $x_t$ evaluated by the logarithmic score 
$$s(x_t, y_t) = - \log(x_t(y_t)),$$
the entropy and divergence are the classical entropy
$$e_{log}(V) = -\E[\log(f_V(V))]$$
and the Kullback-Leibler divergence
$$d_{log} (w,f_V) = - \E \left[ \log \left(\frac{w(V)}{f_V(V)} \right) \right],$$
leading to the following resolution, calibration and decomposition:
$$-\E[\log(X_t(Y_t))] = \underbrace{-\E[\log(f_{Y_t}(Y_t))]}_{UNC} - \underbrace{\E \left[ - \log \left(\frac{f_{Y_t}(Y_t)}{f_{Y_t|X_t}(Y_t)} \right) \right]}_{RES} + \underbrace{\E \left[ - \log \left(\frac{X_t(Y_t)}{f_{Y_t|X_t}(Y_t)} \right) \right]}_{CAL}.$$

For the probabilistic forecasts from table \ref{table_forecasters} and for the logarithmic score the values of the decomposition components are summarized in table \ref{table_distributional_forecasts}, I present some details on the calculations here.

For a normally distributed random variable $V \sim N(\mu_V, \sigma^2_V)$, the entropy is
$$ e_{log}(V) = \frac 1 2 ( \log (2 \pi \sigma^2_V) + 1).$$
The Kullback-Leibler divergence between a normal density forecast $w=f_{N(\mu_w, \sigma^2_w)}$ and the density of $V \sim N(\mu_V, \sigma^2_V)$, $f_V$, is
$$ d_{log} (w,f_V) = \frac 1 2 \left( \frac{(\mu_V - \mu_w)^2}{\sigma^2_w} + \frac{\sigma^2_V}{\sigma^2_w} - \log \left( \frac{\sigma^2_V}{\sigma^2_w} \right) - 1 \right).$$

The expected loss for a normal density forecast $X_t$ of a normal $Y_t$ is
$$\E[- \log(X_t(Y_t))] = \frac 1 2 \left( \log( 2 \pi ) + \E [ \log (\sigma^2_{X_t})] + \E \left[ \left( \frac{Y_t - \mu_{X_t}}{\sigma_{X_t}} \right)^2 \right] \right).$$

Using the formula for the entropy of a normally distributed random variable from above, I get the uncertainty of $Y_t$,
$$e_{log}(Y_t) =  \frac 1 2 ( \log (4 \pi) + 1)$$.

For the unconditional forecaster, the loss equals the uncertainty as calibration and resolution are both zero.

For the informed, the sign-reversed and the noisily informed forecasters, as $\sigma^2_{X_t}=1$, the last term in the loss equation above equals their squared error loss from table \ref{table_mean_forecasts}, thus the expected loss of the former is
$$ MLS = \frac 1 2 ( \log( 2 \pi ) + 1). $$
As he is autocalibrated, his calibration is 0, leading to a resolution of $ RES = \frac 1 2 ( \log(2) )$ by the decomposition formula.

By the argument from above, the accuracy of the sign-reversed forecaster is 
$$ MLS = \frac 1 2 ( \log( 2 \pi ) + 5), $$
its resolution is the same as of the informed forecaster and thus the calibration is $CAL=2$ by the decomposition formula.

For the noisily informed forecaster, the expected score equals
$$ MLS = \frac 1 2 ( \log( 2 \pi ) + 1 + \sigma^2_{\nu}). $$
To calculate calibration and resolution, we have to find the conditional distribution $Y_t|X_t$. Here the fact stated above about the conditional distribution of a random variable given another one that is jointly normal with it can be used again. The first moment has been calculated there and I calculate the second moment now, using in the third step the variance formula from there: 
\begin{align*}
\Var( Y_t | X_t ) &= \Var( \mu_t + \varepsilon_t | \tilde{\mu}_t ) \\
					&= \Var( \mu_t | \tilde{\mu}_t ) +  \Var( \varepsilon_t | \tilde{\mu}_t )\\
					&= \left( 1 - \frac{\Cov(\mu_t, \tilde{\mu}_t)^2}{\Var(\mu_t) \Var(\tilde{\mu}_t)} \right) \Var(\mu_t) +1\\  					
					&= \left( 1 - \frac{\Var(\mu_t)^2}{\Var(\mu_t) \Var(\tilde{\mu}_t)} \right) \Var(\mu_t) +1\\
					&= 2 - \frac{1}{1 + \sigma^2_{\nu}}\\
					&= 1 + \frac{\sigma^2_{\nu}}{1 + \sigma^2_{\nu}}.
\end{align*}

Thus 
$$Y_t|X_t \sim N \left( \frac{1}{1 + \sigma^2_{\nu}} \tilde{\mu}_t,1 + \frac{\sigma^2_{\nu}}{1 + \sigma^2_{\nu}} \right) $$
holds and I use that $X_t = f_{N(\tilde{\mu_t}, 1)}$ to calculate miscalibration:
\begin{align*}
CAL &= \E[ d_{log} (f_{X_t}, f_{Y_t|X_t})]\\
	&= \frac 1 2 \left( \left( \frac{1}{1 + \sigma^2_{\nu}} \tilde{\mu}_t - \tilde{\mu}_t \right)^2 + 1 + \frac{\sigma^2_{\nu}}{1 + \sigma^2_{\nu}} - \log \left( 1 + \frac{\sigma^2_{\nu}}{1 + \sigma^2_{\nu}} \right) - 1 \right)\\
	&= \frac 1 2 \left( \frac{\sigma^4_{\nu}}{1 + \sigma^2_{\nu}} + \frac{\sigma^2_{\nu}}{1 + \sigma^2_{\nu}} - \log \left( 1 + \frac{\sigma^2_{\nu}}{1 + \sigma^2_{\nu}} \right) \right)\\
	&= \frac 1 2 \left( \sigma^2_{\nu} - \log \left( 1 + \frac{\sigma^2_{\nu}}{1 + \sigma^2_{\nu}} \right)  \right).
\end{align*}
For the third equality, note that the expectation of the first summand in the second line is just the calibration from the squared error case that has been calculated above.
 
To calculate resolution, remember that $Y_t \sim N(0, 2)$:
\begin{align*}
RES &= \E[ d_{log} (f_{Y_t}, f_{Y_t|X_t})]\\
&= \E \left[ \frac 1 2 \left( \frac 1 2 \left( \frac{1}{1 + \sigma^2_{\nu}} \tilde{\mu}_t \right)^2 + \frac 1 2 \left( 1 + \frac{\sigma^2_{\nu}}{1 + \sigma^2_{\nu}} \right) - \log \left( \frac 1 2 \left( 1 + \frac{\sigma^2_{\nu}}{1 + \sigma^2_{\nu}} \right) \right) - 1 \right) \right] \\
&= \frac 1 2 \left( \frac 1 2 \frac{1}{1 + \sigma^2_{\nu}} + \frac 1 2 \frac{\sigma^2_{\nu}}{1 + \sigma^2_{\nu}} + \log(2) - \log  \left( 1 + \frac{\sigma^2_{\nu}}{1 + \sigma^2_{\nu}}  \right) - \frac 1 2 \right)\\
&= \frac 1 2 \left( \log(2) - \log  \left( 1 + \frac{\sigma^2_{\nu}}{1 + \sigma^2_{\nu}}  \right) \right).
\end{align*}
Again for the third equality, note that the expectation of the first summand in the second line is just the resolution from the squared error case that has been calculated above.

Now, I turn to the recalibrated forecaster. Note that to construct the recalibrated forecaster, the recalibrated version of the noisily informed forecaster, we set $F^{pr,Rec}_{Y_t} = F_{Y_t|X^{NI}_t}$. This forecaster is autocalibrated and its resolution is the same as the noisily informed forecaster's one, leading to a loss of
$$\frac 1 2 \left( \log( 2 \pi ) + \log \left( 1 + \frac{\sigma_\nu^2}{1 + \sigma_\nu^2} \right) \right)$$
by the decomposition formula.

\section{Robustness Checks and Comparison to Benchmarks for the SPF Forecasts} \label{robustness_plus_benchmarks}

Here, I present two robustness checks and provide results for benchmark forecasts for the SPF inflation forecasts. I do not report the results for GDP growth here as they are very similar. 

\begin{figure}
	\centering
	\includegraphics[width=0.7\linewidth]{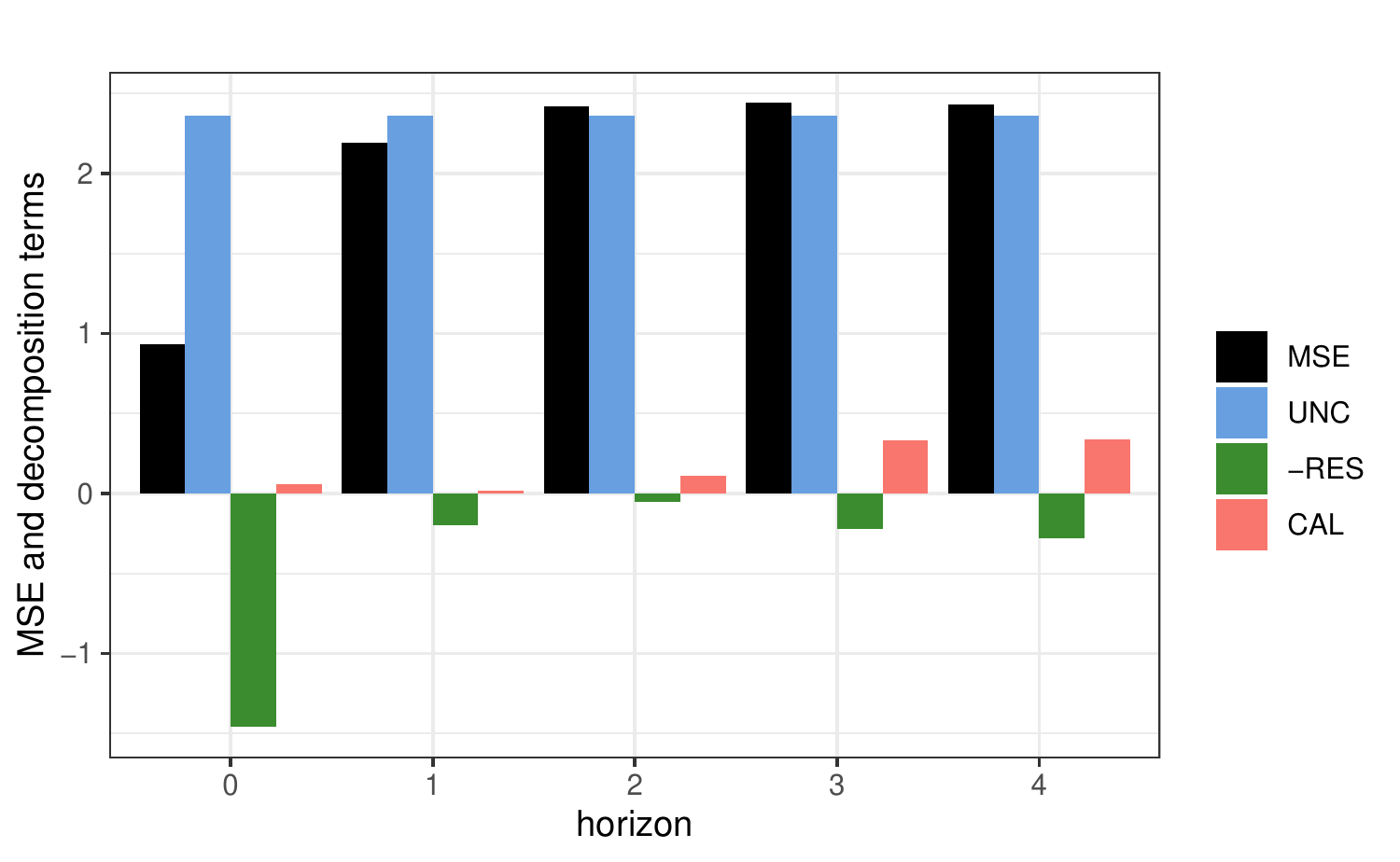}
	\caption{graphical representation of $MSE$ and estimated decomposition terms for quadratic loss for the inflation forecasts of the SPF with the fourth quarter of 2008 removed from the sample}
	\label{barplot_decomposition_SPF_inflation_robust}
\end{figure}

The first robustness check is with respect to the financial crisis, which caused a huge drop in US CPI inflation in the fourth quarter of 2008 as can be seen in figure \ref{timeplot_SPF_inflation}. In figure \ref{barplot_decomposition_SPF_inflation_robust} I present the evaluation results as before in figure \ref{barplot_decomposition_SPF_inflation}, but now with the fourth quarter of 2008 removed from the sample. The observed pattern is the same as before, i.e.\ the high resolution for the forecasts for the current quarter disappears from the first quarter into the future on and miscalibration is almost zero for all horizons. While the outlier thus does not change the ratios of the expected loss and the decomposition terms, it has a strong influence in terms of their size, i.e. it blows them up considerably. 
  
\begin{figure}
	\centering
	\includegraphics[width=0.7\linewidth]{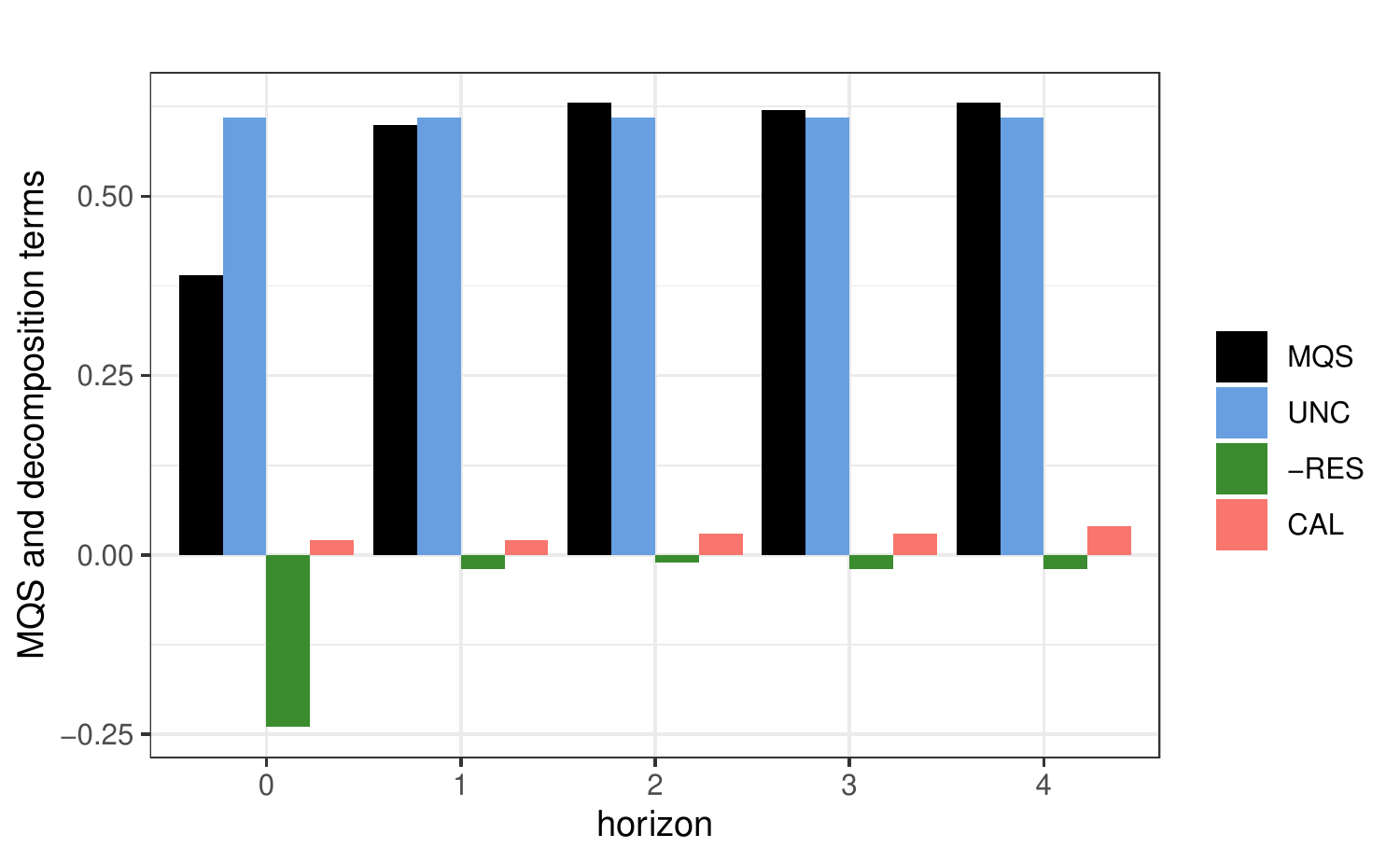}
	\caption{graphical representation of $MQS$ and estimated decomposition terms for the absolute error (i.e.\ the quantile score for $\tau=0.5$) for the inflation forecasts of the SPF}
	\label{barplot_decomposition_SPF_inflation_absoluteerror}
\end{figure}

The second robustness check goes in a very different direction being concerned with the loss function used. As the Philadelphia Fed does not specify a target functional of the forecast distribution or a loss function for the professional forecasters, it is not clear what exactly they forecast. Most often it is assumed that their forecasts are mean forecasts and thus I use the quadratic loss as a consistent scoring function for mean forecasts. However, the forecasters could as well report a different measure of central tendency. For this reason, I also report the evaluation results assuming the absolute error, which is consistent for the median, as a loss function in figure \ref{barplot_decomposition_SPF_inflation_absoluteerror} and the same pattern as in figure \ref{barplot_decomposition_SPF_inflation} emerges again. In principle, the loss function could also be asymmetric and there is a literature trying to estimate the loss function or the functional and then test optimality of the forecasts (see e.g.\ \cite{schmidt2015} and references therein).

\begin{figure}
	\centering
	\includegraphics[width=0.7\linewidth]{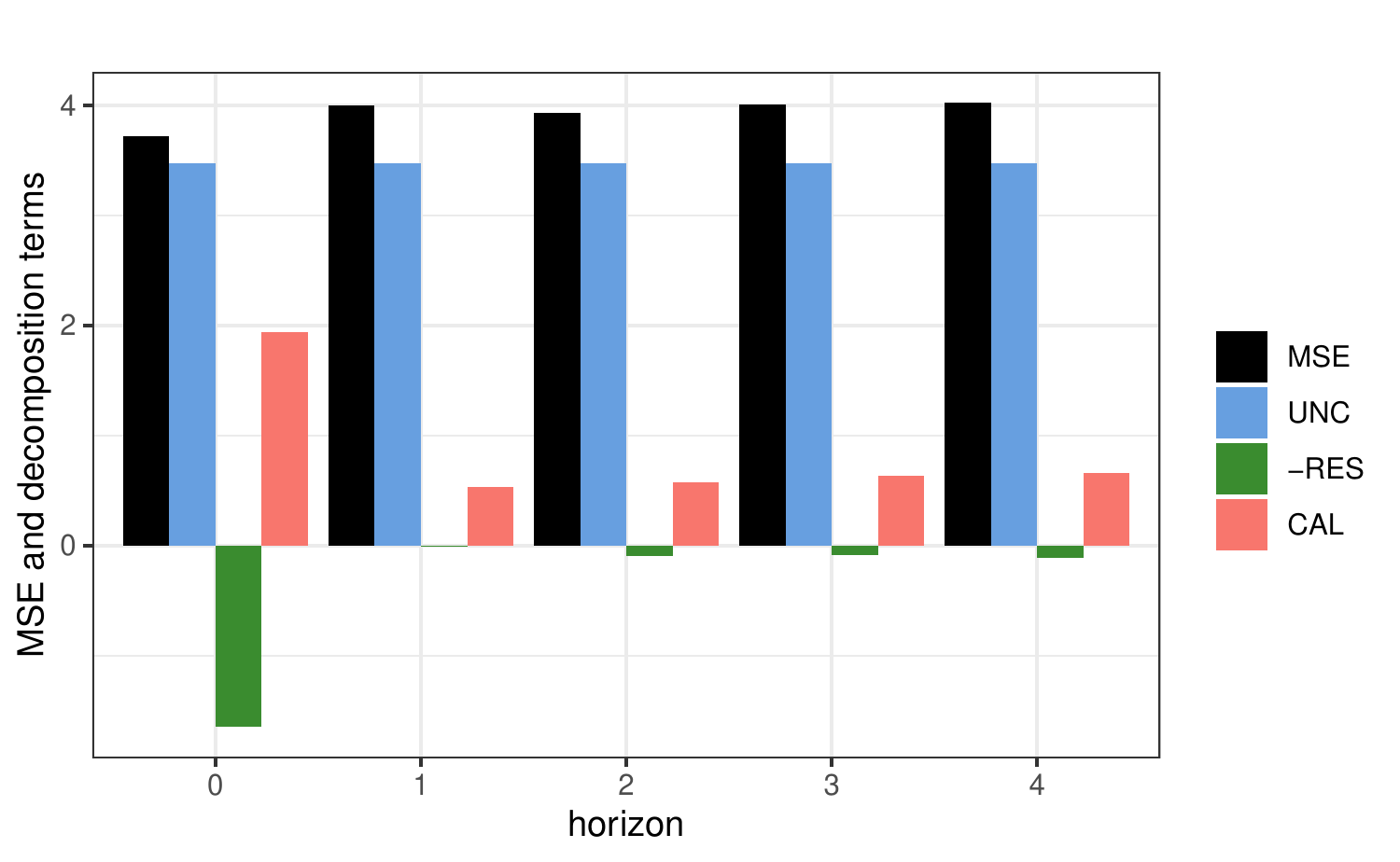}
	\caption{graphical representation of $MSE$ and estimated decomposition terms for quadratic loss for the inflation forecasts obtained by an AR(1) model}
	\label{barplot_decomposition_SPF_inflation_ownARbenchmark}
\end{figure}

A classical benchmark for time series forecasts is the AR model. Thus, I calculated forecasts from an AR(1) model using the Philadelphia Fed's Real-Time Data Set for Macroeconomists\footnote{The real-time data for the CPI were retrieved from https://www.philadelphiafed.org/research-and-data/real-time-center/real-time-data/data-files/cpi at 24.01.2020. As the first vintage available in this dataset is the third quarter of 1994, we used this vintage also for the previous four years needed here.} with an expanding window. The use of real-time data ensures that the model-based forecasts in the pseudo-out-of-sample forecasting exercise use the same information that were available to the professional forecasters at the time of forecasting and not a later-on revised series, see e.g.\ \cite{croushore2001} or \cite{stark2002}, (of course acknowledging that the professional forecasters can use information from the current quarter when they forecast, while these simple model-based forecasts use only information from past quarters). The evaluation results for these forecasts are presented in figure \ref{barplot_decomposition_SPF_inflation_ownARbenchmark}. The forecasts are inferior to the SPF-forecasts as the $MSE$ is bigger than uncertainty for all horizons due to a lack of resolution paired with slight miscalibration. The results for higher-order AR models or for an AR model with the lag order chosen by an information criterion are even worse and are not reported here. The weak performance of the AR models is probably caused by the low levels of autocorrelation in this series of annualized quarter-on-quarter inflation. The Philadelphia Fed also compares its forecasts to AR benchmarks and makes these available for download as well. Using these forecasts I obtain analoguous results.\footnote{These forecasts only start in the third quarter of 1994 due to the availability of the real-time data from then on, but when compared to the SPF-forecasts from then on the same picture as in figure \ref{barplot_decomposition_SPF_inflation_ownARbenchmark} emerges, thus I do not report the exact results here.}

\begin{figure}
	\centering
	\includegraphics[width=0.7\linewidth]{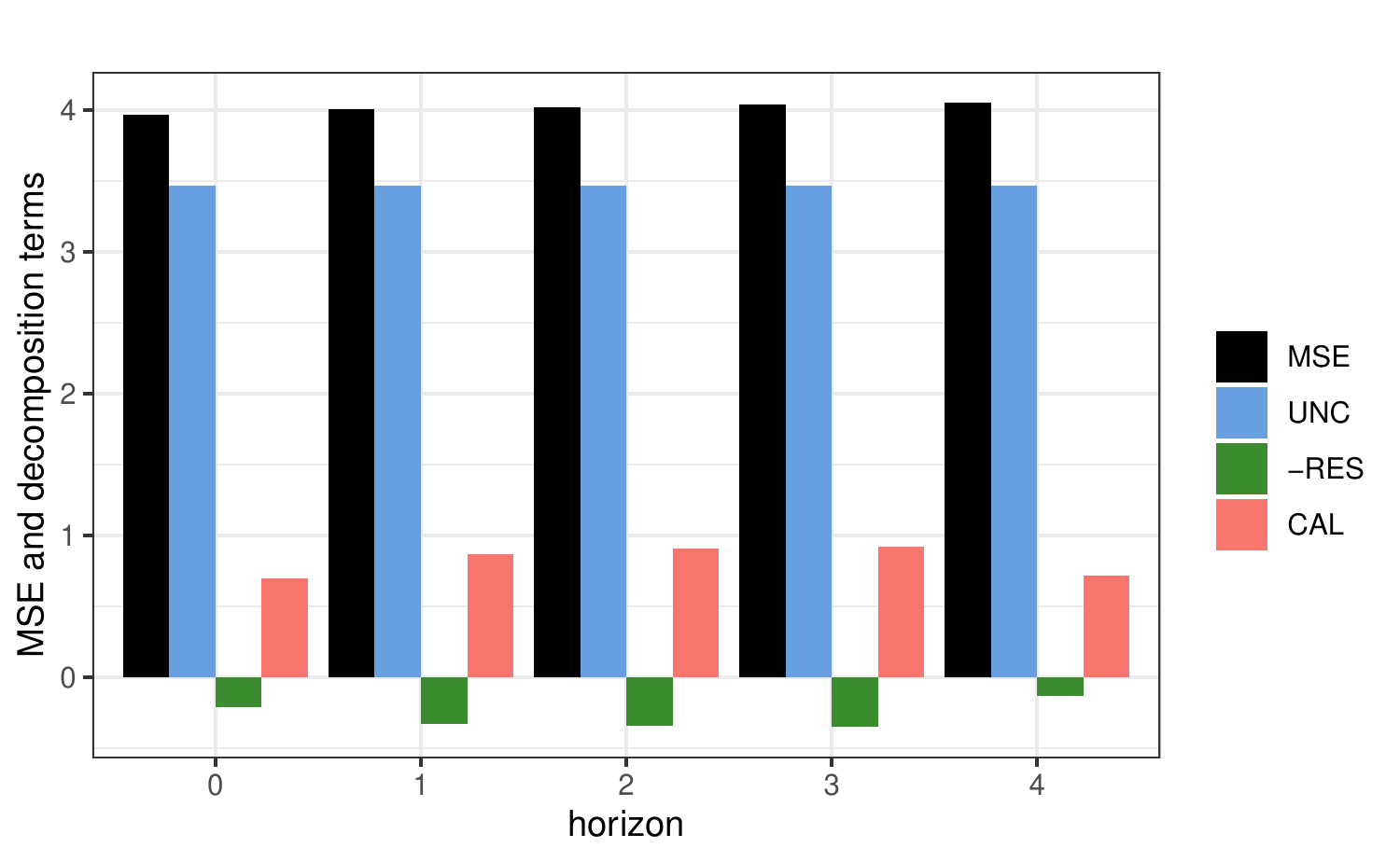}
	\caption{graphical representation of $MSE$ and estimated decomposition terms for quadratic loss for the inflation forecasts obtained by the arithmetic mean of past realizations}
	\label{barplot_decomposition_SPF_inflation_unconditionalbenchmark}
\end{figure}

In figure \ref{barplot_decomposition_SPF_inflation_unconditionalbenchmark} the results for the feasible unconditional benchmark are shown, i.e.\ for forecasts which equal the arithmetic mean of the past of the series. Here I again used the real-time data with expanding window pseudo-out-of-sample forecasting. The $MSE$ is a bit higher than the $MSE$ of the infeasible unconditional benchmark, which is always reported automatically in the Murphy decomposition with the uncertainty component. This is as expected since the estimation of the Murphy decomposition uses the full-sample mean, while the feasible unconditional forecasts only can use the mean of the past data. The performance of this unconditional benchmark is inferior to the performance of the SPF forecasts, but almost identical to the performance of the AR(1) forecasts. 

\section{The Two-Piece Normal Distribution and the BoE's Probabilistic Forecasts} \label{split_normal}

The BoE issues its probabilistic forecasts in the form of a two-piece normal distribution, also known as a binormal or a split normal distribution, see \cite{wallis2014} for its interesting history. The distribution joins two appropriately normalized halves of normal distributions with different variances at their common mode and is thus able to capture skewness. The distribution is governed by three parameters and two popular parametrizations are commonly used (see e.g.\ \cite{julio2006}). One of them is the parametrization used by the BoE (see also \cite{britton1998} or \cite{kotz2004}), where the three parameters are the mode $\mu$, a measure of dispersion $\sigma$, which is not the standard deviation, and an inverse measure of skewness $\gamma$ with $-1 < \gamma < 1$. The density in terms of these three parameters is
$$
f(x) =  
\begin{cases} 
\frac{ A }{ \sqrt{ 2 \pi } \sigma } \exp\left( - \frac{ 1 - \gamma }{ 2 \sigma^2 } ( x - \mu )^2 \right) & \text{if } x \leq \mu \\
\frac{ A }{ \sqrt{ 2 \pi } \sigma } \exp\left( - \frac{ 1 + \gamma }{ 2 \sigma^2 } ( x - \mu )^2 \right)      & \text{if } x > \mu
\end{cases},
$$
with the normalizing constant $A = \frac{ 2 }{ \frac{ 1 }{ \sqrt{ 1 - \gamma } } + \frac{ 1 }{ \sqrt{ 1 + \gamma } } }$.

The BoE publishes in its quarterly inflation reports or more recently in its monetary policy reports the forecasted parameters $\mu$ and $\sigma$, but as a measure of skewness it reports the difference between the mean and the mode of the distribution $\xi$. The parameter $\gamma$ can be calculated from the three reported parameters by
$$\gamma = \begin{cases} 
-\sqrt{ 1 - \left( \frac{ \sqrt{ 1 + 2 \beta } - 1  }{ \beta } \right) ^2 }& \text{if } \xi > 0 \\
0 & \text{if } \xi = 0 \\
\sqrt{ 1 - \left( \frac{ \sqrt{ 1 + 2 \beta } - 1  }{ \beta } \right) ^2 } & \text{if } \xi < 0
\end{cases},$$
where $\beta = \frac{ \pi }{ 2 \sigma^2} \xi^2$.\footnote{I spell out the used parametrization and the formula for $\gamma$ explicitly to ensure reproducibility of my results. In this case this is especially important as this formula is often reported with the wrong sign and containing an undefined expression for $\xi=0$  in the literature.}

For the formula for the quantiles of this distribution see e.g.\ \cite{julio2006}.

\end{document}